\documentclass[twocolumn]{emulateapj}  
\usepackage{csvsimple}
\usepackage{arydshln}
\usepackage{epstopdf}
\usepackage{ulem}
\usepackage{amssymb}
\usepackage{times}
\usepackage{lineno}
\usepackage{graphicx}
\usepackage[usenames,dvipsnames]{pstricks}
\usepackage{psfrag}
\usepackage{lipsum}
\usepackage{natbib}
\usepackage[colorlinks=true,linkcolor=blue,citecolor=blue]{hyperref}
\def\aj{AJ}
\def\apj{ApJ}
\def\apjl{ApJ}
\def\apjs{ApJS}
\def\aap{A\&A}
\def\mnras{MNRAS}
\def\physrep{Phys.~Rep.}

\newcommand{\be}{\begin{equation}}
\newcommand{\ee}{\end{equation}}
\newcommand{\bary}{\begin{eqnarray}}
\newcommand{\eary}{\end{eqnarray}}

\begin{document}

%
\title{Modeling the observations of GRB 180720B:  From radio to sub-TeV gamma-rays}
%
\author{N.~ Fraija\altaffilmark{1$\dagger$}, S. Dichiara\altaffilmark{2,3},     A.C. Caligula do E. S. Pedreira\altaffilmark{1}, A.~ Galvan-Gamez\altaffilmark{1}, R. L. Becerra\altaffilmark{1}, A. Montalvo\altaffilmark{1}, J. Montero\altaffilmark{1},  B. Betancourt Kamenetskaia\altaffilmark{1} and B.B. Zhang\altaffilmark{4,5}}
\affil{$^1$Instituto de Astronom\'ia, Universidad Nacional Aut\'{o}noma de M\'{e}xico, Apdo. Postal 70-264, Cd. Universitaria, Ciudad de M\'{e}xico 04510}
\affil{$^2$ Department of Astronomy, University of Maryland, College Park, MD 20742-4111, USA}
\affil{$^3$ Astrophysics Science Division, NASA Goddard Space Flight Center, 8800 Greenbelt Rd, Greenbelt, MD 20771, USA}
\affil{$^4$ School of Astronomy and Space Science, Nanjing University, Nanjing 210093, China}
\affil{$^5$ Key Laboratory of Modern Astronomy and Astrophysics (Nanjing University), Ministry of Education, China}
\email[$\dagger$ ]{nifraija@astro.unam.mx}
\begin{abstract}
Early and late multiwavelength observations play an important role in determining the nature of the progenitor, circumburst medium,  physical processes and  emitting regions associated to  the spectral and temporal features of bursts.   GRB 180720B is a long and powerful burst detected by a large number of observatories in multiwavelenths that range from radio bands to sub-TeV gamma-rays.  The  simultaneous  multiwavelength observations were presented over multiple  periods of time beginning just after the trigger time and extending for more than 30 days.  The temporal and spectral analysis of Fermi LAT observations suggests that it presents similar characteristics to other bursts detected by this instrument.   Coupled with X-ray and optical observations,  the standard external-shock model in a homogeneous medium is favored by this analysis.   The X-ray flare is consistent with the synchrotron self-Compton (SSC) model from the reverse-shock region evolving in a thin shell and long-lived LAT, X-ray and optical data with the standard synchrotron forward-shock model.  The best-fit parameters derived with the Markov chain Monte Carlo simulations indicate that the outflow is endowed with magnetic fields and that the radio observations are in the self-absorption regime.   The SSC forward-shock model with our parameters can explain  the LAT photons beyond the synchrotron limit as well as the emission recently reported by the HESS Collaboration.
\end{abstract}
%
\keywords{Gamma-rays bursts: individual (GRB 180720B)  --- Physical data and processes: acceleration of particles  --- Physical data and processes: radiation mechanism: nonthermal --- ISM: general - magnetic fields}

\section{Introduction}
The most energetic gamma-ray sources in the observable universe are gamma-ray bursts (GRBs).   These events display short and bright irregular flashes of gamma-rays  originated inside the relativistic outflows launched by a central engine.  This engine may result from a merger of either two neutron stars (NSs) or a NS and a black hole (BH), in which case the events are known as ``short GRBs (sGRBs)".  On the other hand, if the engine comes from a cataclysmic event at the end of the life cycle of massive stars, these events are referred to as ``long GRBs (lGRBs)". The duration of  sGRBs lasts  $\lesssim$ few seconds and lGRBs last $\gtrsim$ few seconds \cite[see, i.e.][for reviews]{2004IJMPA..19.2385Z, 2015PhR...561....1K}.   The most accepted mechanism for producing the bright flashes known as the ``prompt emission"  is the standard fireball model \citep{1992MNRAS.258P..41R, 1997ApJ...476..232M}.  According to this  model, a long-lasting ``afterglow" emission in wavelengths ranging from radio bands to gamma-rays is also expected.  The prompt emission is expected when  inhomogeneities in the jet lead to internal collisionless shocks \citep[when matter ejected with low velocity is hit by matter with high velocity;][]{1994ApJ...430L..93R}  and the afterglow when the relativistic outflow sweeps up enough external ``circumburst" medium \citep{1997ApJ...476..232M}. The transition between the prompt and early afterglow emission is determined by the steep decay usually interpreted  as the high-latitude emission \citep{2000ApJ...541L..51K, 2006ApJ...642..389N} and by an X-ray flare or optical flash explained in terms of the reverse shock \citep{2007ApJ...655..973K, 2007ApJ...655..391K, 2000ApJ...545..807K, 2018ApJ...859...70F, 2019arXiv190405987B}.\\   
Multiwavelength observations play an important role in determining the physical processes and emitting places associated with the spectral and temporal features of bursts \citep{2013ApJ...763...71A, 2015ApJ...804..105F, 2017ApJ...848...15F}.  The early-time afterglow observations are useful to determine the nature of the central engine and constrain the density of the circumburst medium \citep{2016ApJ...818..190F, 2016ApJ...831...22F, 2017ApJ...837..116B, 2019ApJ...872..118B}.  In these cases, GRBs become potentially more interesting and informative, allowing afterglow models to be tested more rigorously.\\
Since the discovery of the first GRB in 1967 by Vela satellites \citep{1973ApJ...182L..85K}, the detection of high-energy (HE) photons  ($\gtrsim 100\,{\rm MeV}$) has been possible in only a small fraction of them ($\sim$ 150 bursts\footnote{https://fermi.gsfc.nasa.gov/ssc/observations/types/grbs/lat\_grbs/}).   At higher energies, in the GeV energy range, few detections have been reported and interpreted in the leptonic and hadronic scenarios operating at several possible emitting regions.  The HE and very-high-energy  (VHE; $\gtrsim 10\, {\rm GeV}$)  photons have been detected during the prompt and long-lived emission \citep{2019ApJ...878...52A}.  Different analyses of multiwavelength observations covering from radio to GeV energies have indicated that the HE and VHE emission is produced during  the internal and external shocks \citep[e.g., see][]{2015PhR...561....1K}.  During the afterglow phase the synchrotron emission from electrons accelerated in the external shocks dominates from radio wavelengths to  gamma-rays, and the synchrotron self-Compton (SSC) emission and photo-hadronic processes \citep{2000ApJ...541L...5M, 2004ApJ...604L..85A, 2014MNRAS.437.2187F}  are expected to dominate in the GeV - TeV energy range \citep{2001ApJ...559..110Z}.    The maximum photon energy radiated by the synchrotron process during the deceleration phase is $\sim 10~{\rm GeV}~\left(\frac{\Gamma}{100}\right)\left(1+z\right)^{-1}$, where $\Gamma$ is the bulk Lorentz factor and $z$ is the redshift  \citep{2010ApJ...718L..63P, 2009ApJ...706L.138A, 2011MNRAS.412..522B}.   Consequently,  we accentuate that the VHE photons below the maximum photon energy radiated in the  synchrotron forward-shock model can be interpreted in this scenario, but beyond the synchrotron limit other scenarios must be invoked to explain them.\\
GRB 180720B was detected and followed up by the three instruments onboard the Swift satellite  \citep{2018GCN.22973....1P, 2018GCN.22998....1B};  the Burst Area Telescope (BAT),  the X-ray Telescope (XRT) and the Ultra-Violet/Optical Telescope (UVOT), by both instruments onboard the Fermi satellite \citep{2018GCN.22981....1R, 2018GCN.22980....1B}, the Gamma-ray Burst Monitor (GBM) and the Large Area Telescope (LAT), by the MAXI Gas Slit Camera (GSC) \citep{2018GCN.22993....1N}, by Konus-Wind \citep{2018GCN.23011....1F}, by the Nuclear Spectroscopic Telescope Array (NuSTAR) \citep{2018GCN.23041....1B}, by the CALorimetric Electron Telescope (CALET) Gamma-ray Monitor \citep{2018GCN.23042....1C}, by the Giant Metrewave Radio Telescope \citep[GMRT;][]{2018GCN.23073....1C},  by the Arcminute Microkelvin Imager Large Array \citep[AMI-LA;][]{2018GCN.23037....1S} and by several optical ground telescopes \citep{2018GCN.23040....1I, 2018GCN.23033....1Z, 2018GCN.23024....1J, 2018GCN.23023....1L, 2018GCN.23021....1C, 2018GCN.23020....1S, 2018GCN.23017....1W, 2018GCN.23004....1H}.\\
In this paper,  we derive and analyze the LAT light curve and spectrum for GRB 180720B and show that it exhibits similar features to other powerful bursts.  We show that the photon flux light curve recently reported in the second GRB catalog \citep{2019ApJ...878...52A} is consistent with the one obtained in this work. In addition, we determine the GBM light curve and show that it is consistent with the prompt emission.   Analyzing the multiwavelength observations covering from radio bands to GeV gamma-rays, we show that LAT,  X-ray, optical and radio observations are consistent  with the synchrotron forward-shock model in a homogeneous medium. We also show that the LAT photons beyond the synchrotron limit as well as the emission recently reported by the HESS Collaboration are consistent with the SSC forward-shock model.  The X-ray flare is consistent with SSC emission from the reverse-shock region in a homogeneous medium.   The paper is arranged as follows. In Section 2 we present multiwavelength observations and/or data reduction. In Section 3 we describe the multiwavelength observations through the synchrotron forward-shock model and the SSC reverse-shock model. In Section 4, we exhibit the discussion and results of the analysis done using the multiwavelength data.   Finally,  in Section 5 we give a brief summary and emphasize our conclusions.  The convention $Q_{\rm x}=Q/10^{\rm x}$  in cgs units will be adopted throughout this paper.  The sub-indexes ``f" an ``r" are related to the derived quantities in the forward and reverse shocks, respectively.
\vspace{2cm}
\section{Observations and Data Analysis }\label{sec:observations}
\subsection{Fermi LAT data}
\begin{figure}[h!]
{\centering
\resizebox*{\textwidth}{0.6\textheight}
{\includegraphics{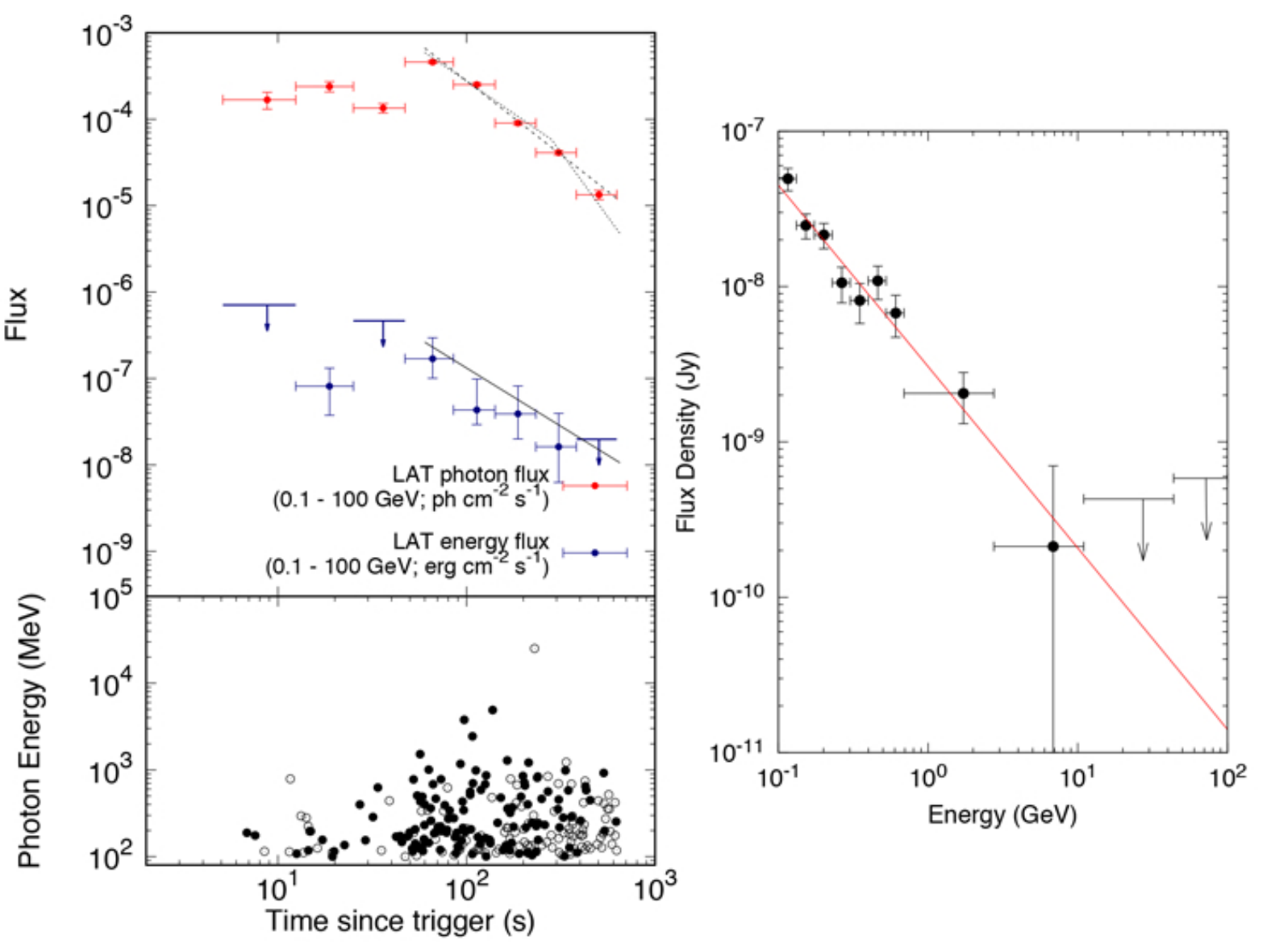}}
}
\caption{The left-hand panel shows the Fermi-LAT energy flux (blue) and photon flux (red) light curves obtained between 0.1 and 300 GeV (upper panel) and all the photons with energies $> 100$ MeV in the direction of  GRB 180720B (lower panel).  The filled circles correspond to the individual photons with a $>0.9$ probability of being associated with this burst and the open circles indicate the LAT gamma Transient class photons. The right-hand panel shows the Fermi LAT spectrum  obtained between 0.15 and 900.45 s.}
\label{LAT}
\end{figure}

%
%
The data files used for this analysis were obtained from the online data website.\footnote{https://fermi.gsfc.nasa.gov/cgi-bin/ssc/LAT/LATDataQuery.cgi}  They contain information 600 s before up to 1000 s from the trigger time ($T_0$) \cite[2018-07-20 14:21:39.65\,  UTC;][]{2018GCN.22980....1B}. Fermi-LAT data was analyzed in the 0.1-100 GeV energy range and the time interval of $T_0+10\,{\rm s}$ up to $T_0+630\,{\rm s}$ with the Fermi Science tools\footnote{https://fermi.gsfc.nasa.gov/ssc/data/analysis/software/} \texttt{ScienceTools v10r00p05}.  For this analysis we adopt the \texttt{P8R2\_TRANSIENT020\_V6} response,  following the unbinned likelihood analysis presented by the Fermi-LAT team.\footnote{https://fermi.gsfc.nasa.gov/ssc/data/analysis/scitools/likelihood\_tutorial.html}  Using the \texttt{gtselect} tool, we select, with a eventclass 16, a region of interest (ROI) around the position of this burst within a radius of 10$^{\circ}$.  We apply a cut on the zenith angle above 100$^{\circ}$.   Then, we  select the appropriate time intervals (GTIs)  using the \texttt{gtmktime} tool on the data selected before considering the ROI cut. In order to define the model needed to describe the source and the diffuse components, 
we use \texttt{modeleditor}. We define a point source at the position of this burst, assuming a power-law spectrum, and we define a galactic diffuse component using GALPROP \texttt{gll\_iem\_v06} as well as the extragalactic background \texttt{iso\_P8R3\_SOURCE\_V2}.\footnote{https://fermi.gsfc.nasa.gov/ssc/data/access/lat/BackgroundModels.html} We use \texttt{gtdiffrsp} to take into account all of these components. Following the likelihood procedure, we produce a lifetime cube with the tool \texttt{gtltcube}, using a step $\delta \theta=0.025$, a bin size of 0.5 and a maximum zenith angle of 100$^{\circ}$. The exposure map was created using \texttt{gtexpmap}, considering a region of 30$^{\circ}$ around the GRB position and defining 100 spatial bins in longitude/latitude and 50 energy bins. Finally, we perform the likelihood analysis with \texttt{gtlike} obtaining a photon flux of (5.2$\pm$0.4$)\times$10$^{-5}$ photons cm$^{-2}$ s$^{-1}$ and a test statistic of 883.267. \\
We obtain the photons with a probability greater than 90$\%$ to be associated to GRB 180720B with the \texttt{gtsrcprob}. In this case, we use \texttt{gtbin} in order to obtain the light curve and considering 8 logarithmically uniform temporal bins.  The photon flux is generated by using the counts and the exposure in each bin. The exposure is obtained with \texttt{gtexposure}.  To derive the energy flux we compute the spectra integrated over each interval assuming logarithmic binning for the energy between 100 MeV and 100 GeV. Then we obtain the Detector Response Matrix with the \texttt{gtrspgen} tool assuming a point like source, a  maximum cutoff angle of 60$^{\circ}$ and a bin size of 0.05 into 30 logarithmically uniform bins between 100 and 100 GeV.  Finally, we derive the backgound spectra using \texttt{gtbkg} and subtract it to the source using XSPEC (v12.10.1; Arnaud 1996) in order to obtain the energy flux with 90\% confidence error.\\
%
%
%
The left-hand panel of Figure \ref{LAT} displays the Fermi LAT energy flux (blue) and photon flux (red) light curves  (upper panel) and all the photons with energies $> 100$ MeV associated with GRB 180720B (lower panel).  The filled circles in the bottom panel correspond to the individual photons and their energies with a $>0.9$ probability of being associated with GRB 180720B and the open circles indicate the LAT gamma Transient class photons. Dotted and dashed lines on the photon flux light curve correspond to the best-fit curves using a power-law (PL) and a broken power-law (BPL) function. The best-fit values are $\alpha=1.81\pm 0.16$ for the PL function, and $\alpha_1=1.49\pm 0.12$ and $^*\alpha_2=3.09\pm 0.64$ for the BPL function. These values are  consistent with the ones from the photon flux light curve reported in Table 5 of \cite{2019ApJ...878...52A}.  The right-hand panel of Figure \ref{LAT} shows the Fermi LAT spectrum.\\
We modeled the energy flux light curve  and spectrum using the closure relation $F^{\rm syn}_{\rm \nu, f}\propto t^{-\alpha_{\rm LAT}} \epsilon^{-\beta_{\rm LAT}}_\gamma$.  The best-fit values of the temporal and spectral PL indexes were $\alpha_{\rm LAT}=1.45\pm0.53$ ($\chi^2$ = 1.11) and $\beta_{\rm LAT}=1.17\pm0.15$ ($\chi^2$ = 1.09), respectively.  These PL indexes are compatible with the third PL segment  of the synchrotron forward-shock model ($\propto t^{-\frac{3p-2}{4}} \epsilon_\gamma^{-\frac{p}{2}}$) for $p\approx 2.6\pm0.2$. It is worth emphasizing that this PL segment is equal for the wind and homogeneous afterglow model.\\
\\
Some relevant characteristics can be observed in the lower panel of Fig. \ref{LAT}: i) the first HE photon (101  MeV) was detected 19.4 s after the trigger time,   ii) this burst exhibited 130 photons with energy greater than 100 MeV and 8 photons with energies greater than 1 GeV,   iii) the highest-energy photon\footnote{This photon was associated to this burst with a probability of 1.} exhibited in the LAT observations (4.9 GeV) was detected 142.43 s after the trigger time and iv) the photon density increased dramatically for a time longer than $\gtrsim$ 50 s.
\subsection{Fermi GBM data}
\begin{figure}[h!]
{\centering
\resizebox*{0.45\textwidth}{0.33\textheight}
{\includegraphics{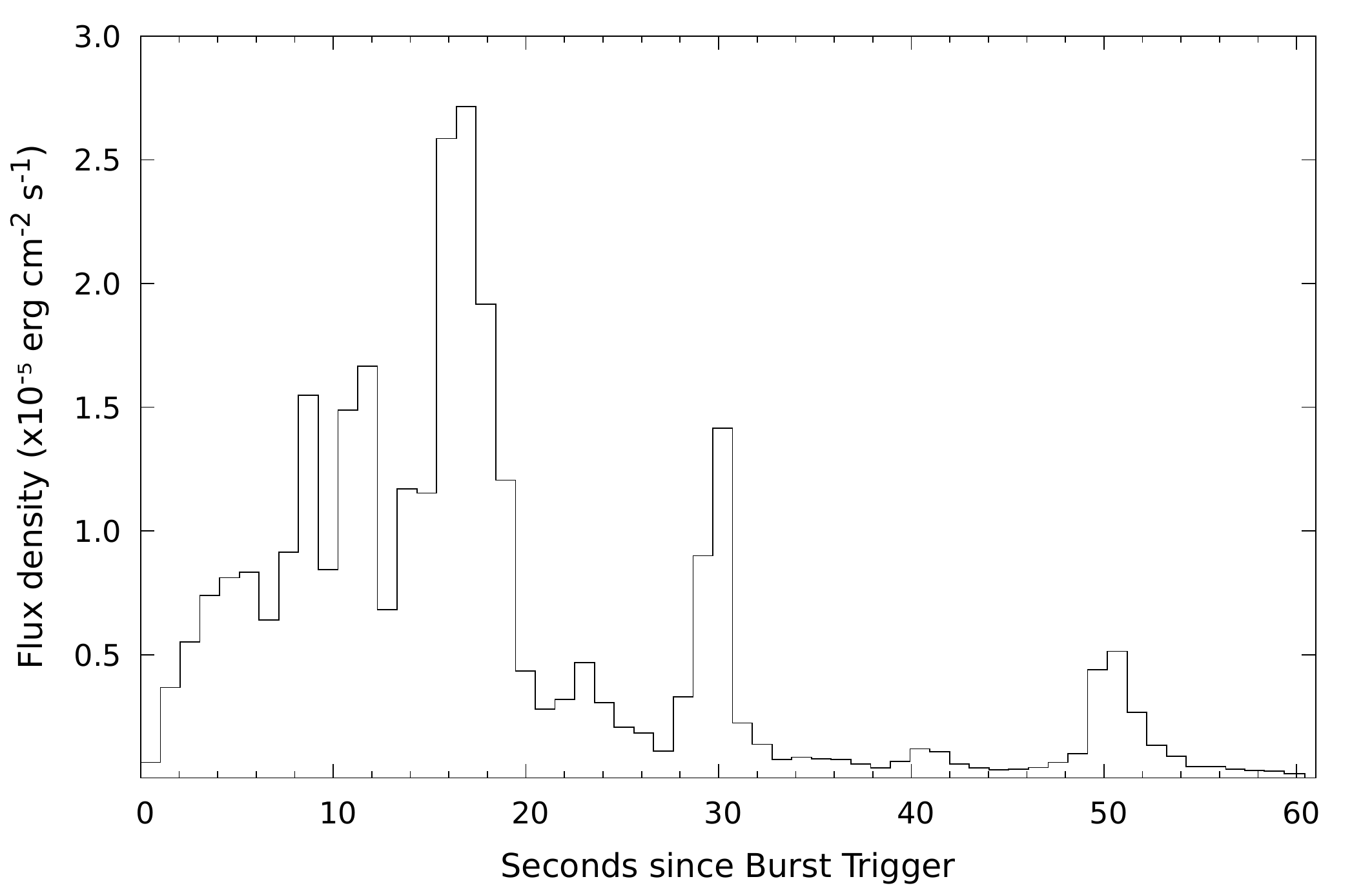}}
\resizebox*{0.45\textwidth}{0.33\textheight}
{\includegraphics{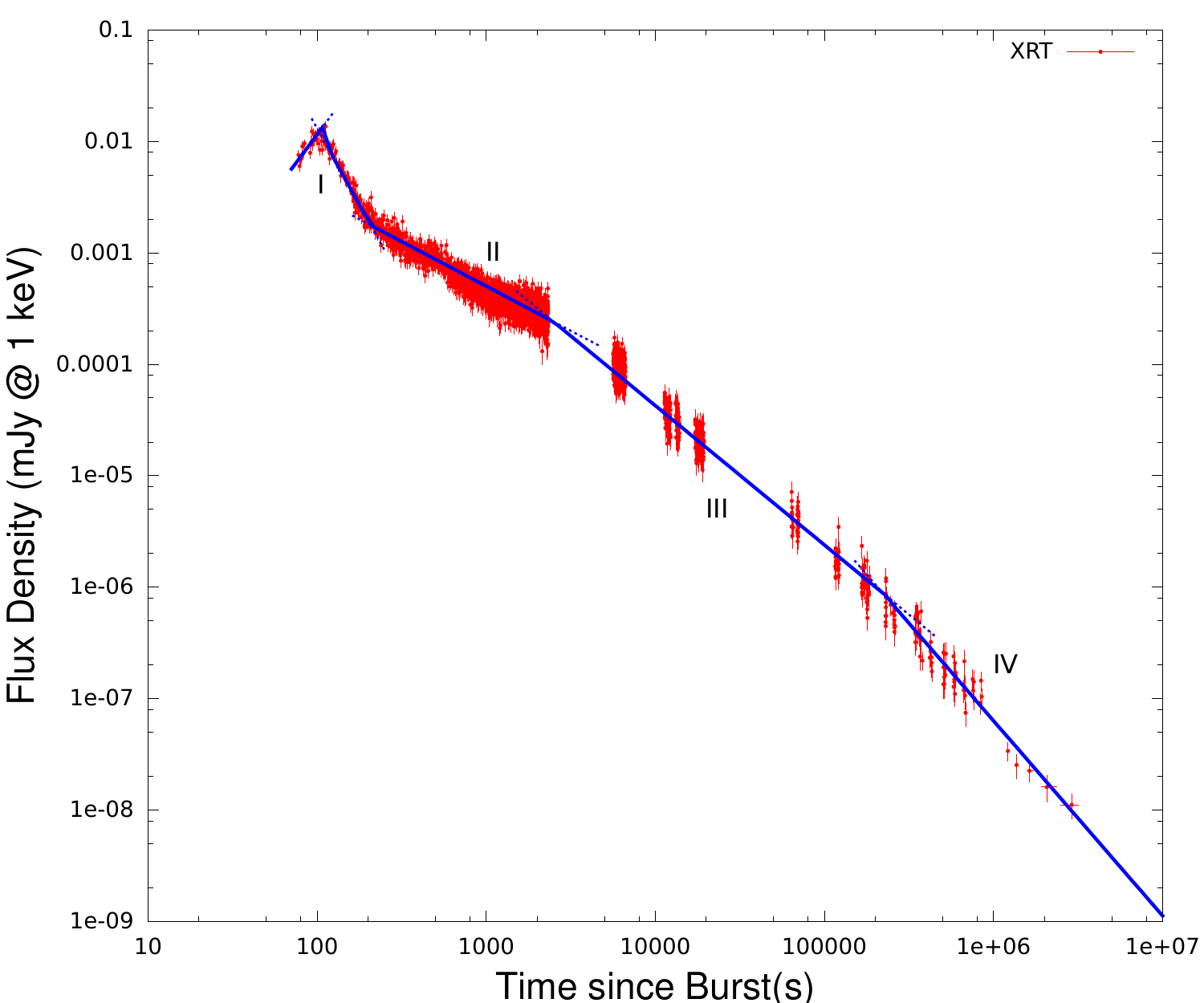}}
\resizebox*{0.70\textwidth}{0.28\textheight}
{\includegraphics{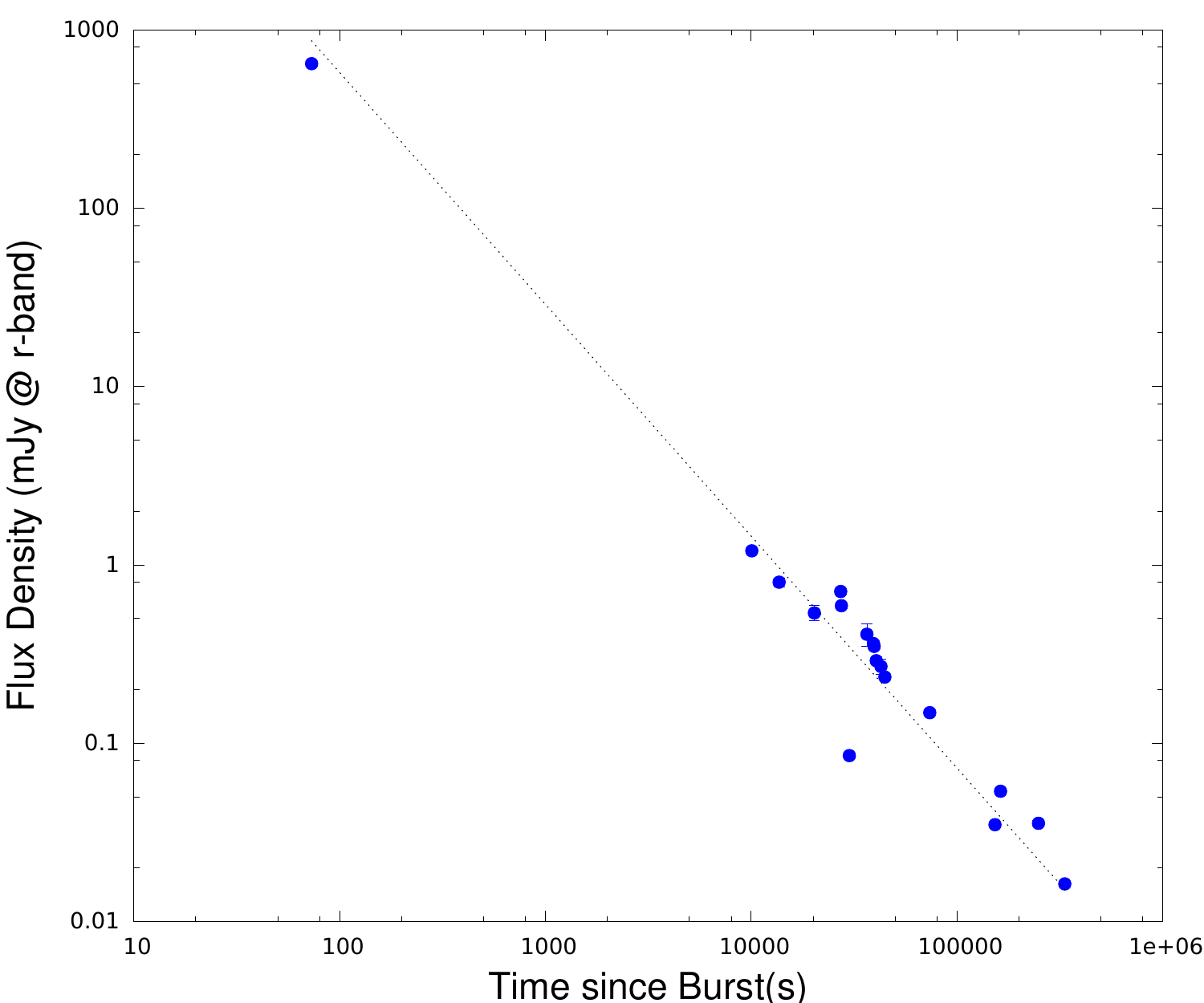}}\\
}
\caption{The upper left-hand panel shows the GBM light curve obtained in the 10 - 1000 keV energy range. GBM data  were reduced using the public database at the Fermi website. The upper right-hand panel shows the X-ray light curve obtained with the Swift XRT  instrument at 1 keV. Blue lines correspond to the best-fit curves using PL functions. The Swift data were obtained using the  publicly available database at the official Swift web site.  The lower panel shows the optical R-band light curve with the best-fit PL function.}
\label{LCs}
\end{figure}

The Fermi-GBM data was obtained using the public database at the GBM website.\footnote{http://fermi.gsfc.nasa.gov/ssc/data} The event data files were obtained using the Fermi GBM Burst Catalog\footnote{https://heasarc.gsfc.nasa.gov/W3Browse/fermi/fermigbrst.html} and the GBM trigger time for GRB 180720B at 14:21:39.65 UT \citep{2018GCN.22981....1R}.   Flux values were derived using  the spectral analysis package Rmfit version 432.\footnote{https://fermi.gsfc.nasa.gov/ssc/data/analysis/rmfit/}  In order to analyze the signal we used the time tagged event (TTE) files of the two triggered NaI detectors $n_{7}$ and $n_{11}$ and the BGO detector $b_{1}$. Two different models were used to fit the spectrum in the energy range of 10 - 1000 keV over different time intervals.  The Band  and the Comptonized models were used to fit the spectrum during the time interval [0.000, 60.416 s].  Each time bin was chosen by adopting the minimum resolution required to preserve the shape of the time resolution. \\
 \\
The upper left-hand panel in Figure \ref{LCs} displays the GBM light curve in the 10 - 1000 keV energy range.   This light curve shows  a bright, fast-rise exponential-decay (FRED)-like peak with a maximum flux of $2.74\times 10^{-5}\,{\rm erg\,cm^{-2}\,s^{-1}}$ at 15 s, followed by  two significant peaks with fluxes of  $1.64\times 10^{-5}$ and $5.5\times 10^{-6} \,\,{\rm erg\,cm^{-2}\,s^{-1}}$ at 26 s and 50 s, respectively. The fluence over the prompt emission was $(2.985\pm 0.001)\times 10^{-4}\,{\rm erg\, cm^{-2}}$ which corresponds to an equivalent isotropic energy of $3\times10^{53}\,{\rm erg}$ for a measured redshift  of $z=0.654$ \citep{2018GCN.22996....1V}. This light curve exhibits a high-variability $\delta t/t \ll 1$,\footnote{$\delta t$ is the width of the peak and $t$ is the timescale of the flux} which favors the prompt phase scenario. Theoretically, this timescale is interpreted as the time difference of two photons emitted at two different radii \citep{1997ApJ...485..270S}.
\vspace{1cm}
\subsection{X-ray data}
The Swift BAT triggered on GRB 180720B on July 20, 2018 at 14:21:44 UT.   This instrument located this burst  at the coordinates: ${\rm R.A.=00h 02m 07s}\,$ and ${\rm Dec}= -02d 56' 00''\, (J2000)$ with an uncertainty  of 3 arcmin.  The XRT instrument started observing this burst 86.5 s after the trigger, and monitored the afterglow for the following 33.5 days  for a total net exposure of  13 ks in Windowed Timing (WT) mode and $2.8\times10^3\, {\rm ks}$ in Photon Counting (PC) mode.  The Swift data used in this analysis are publicly available in the website database.\footnote{http://www.swift.ac.uk/xrtproducts/}  In the WT mode, the reported value of the photon spectral index was $\Gamma_{\rm X}=\beta_{\rm X}+1=1.761\pm0.01$ for a galactic (intrinsic) absorption of  $N_{H}= 3.92 (17.7\pm 0.7)\times 10^{20}\,{\rm cm^{-2}}$.  In the PC mode, the reported value of the photon spectral index was $\Gamma_{\rm X}=\beta_{\rm X}+1=1.83\pm0.06$ for  $N_{H}= 3.92 (24.0\pm 0.4)\times 10^{20}\,{\rm cm^{-2}}$.\\
The upper right-hand panel in Figure \ref{LCs} shows the Swift X-ray light curve obtained with  the XRT (WC and PC modes) instrument at 1 keV.  The flux density of the XRT data was extrapolated from 10  keV to 1 keV using the conversion factor introduced in \cite{2010A&A...519A.102E}. The blue curves correspond to the best-fit  PL functions obtained using the chi-square minimization algorithm installed in ROOT \citep{1997NIMPA.389...81B}. In accordance with  the observational X-ray data,  three PL segments  ($t^{-\alpha_{\rm X}}$)  with an X-ray flare were identified in this light curve.   We evaluated the X-ray light curve at
four time intervals, designated as epochs I, II, III and IV: $70 \lesssim t\lesssim 200\,{\rm s}$ (I),  $200 \lesssim t\lesssim 2.5\times 10^3\,{\rm s}$ (II), $2500 \lesssim t\lesssim 2.6\times 10^5\,{\rm s}$ (III) and $t\geq 2.6\times 10^5$ (IV). The time intervals were chosen in accordance with the variations of each slope.   The temporal PL indexes are $\alpha_{\rm X, rise}=-2.05\pm0.27 (\chi^2/ndf=1.12)$ and $\alpha_{\rm X, decay}=2.74\pm0.08\, (1.27)$ during epoch ``I" and $\alpha_{\rm X}=0.79\pm 0.06\,(1.31)$, $1.26\pm 0.06\,(1.29)$ and $1.75\pm0.09\,(1.21)$ for epochs ``II", ``III" and ``IV", respectively.  The best-fit values of each epoch are reported in Table \ref{table1}. 
%
%
\subsection{Optical data}
GRB 180720B started to be detected in the optical and near infrared (NIR) bands on July 20, 2018 at 14:22:57 UT, 73 s after the trigger time \citep{2018GCN.22977....1S}. Using the HOWPol and HONIR instruments attached to the 1.5-m Kanata telescope, these authors reported a bright optical R-band counterpart  of $m_R=9.4$ mag.  \cite{2018GCN.22996....1V} observed the optical counterpart of this burst using the VLT/X-shooter spectrograph. They detected a bright continuum with some absorption lines (Fe II,  Mg II, Mg I and Ca II) associated to  a redshift of  $z=0.654$. Additional photometry in different optical bands is reported in \cite{2018GCN.22976....1M, 2018GCN.22979....1R, 2018GCN.22983....1I, 2018GCN.22988....1C, 2018GCN.23004....1H, 2018GCN.23017....1W, 2018GCN.23020....1S, 2018GCN.23023....1L}.\\
The lower panel in Figure  \ref{LCs} shows the  optical light curve of GRB 180720B in the  R-band. The solid line represents the best-fit PL function.    Optical data taken from the GCN circulars  reported in this subsection were detected by different telescopes.  The optical observations  with  their uncertainties were obtained using the standard conversion for AB magnitudes shown in \cite{1996AJ....111.1748F}.  The optical data were corrected by the galactic extinction using the relation derived in \cite{2019ApJ...872..118B}. The values of $\beta_{\rm O}=0.80\pm 0.04$ for optical filters and a reddening of $E_{B-V}=0.037$ mag \citep{2019GCN.23702....1B} were used.    The best-fit value of the temporal decay is  $1.22\pm0.02$ ($\chi^2/ndf=1.05$; see Table \ref{table2}). \\
\subsection{Radio data}
\cite{2018GCN.23037....1S} observed the position of this burst with AMI-LA at 15.5 GHz for 3.9 hours.  The observations began  2 days after the BAT trigger, providing an integrated flux of $\sim$ 1 mJy.   \cite{2018GCN.23073....1C} detected GRB 180720B with GMRT at the 1.4 GHz band, reporting a flux of $\sim 390\pm 59\, {\rm \mu Jy}$.
\subsection{HESS observations}
During the Cherenkov Telescope Array (CTA) Science Symposium 2019, the HESS collaboration reported the discovery of late-time VHE emission from GRB 180720B. The VHE emission with $\sim$ 5 $\sigma$  was in the energy range from 100 to 400 GeV. The observations began $\sim$ 10 hours after the burst trigger.\footnote{https://indico.cta-observatory.org/event/1946/timetable/}
%
%
%
%
\section{Broadband Afterglow Modeling}
\begin{figure}[h!]
{\centering
\resizebox*{0.49\textwidth}{0.36\textheight}
{\includegraphics{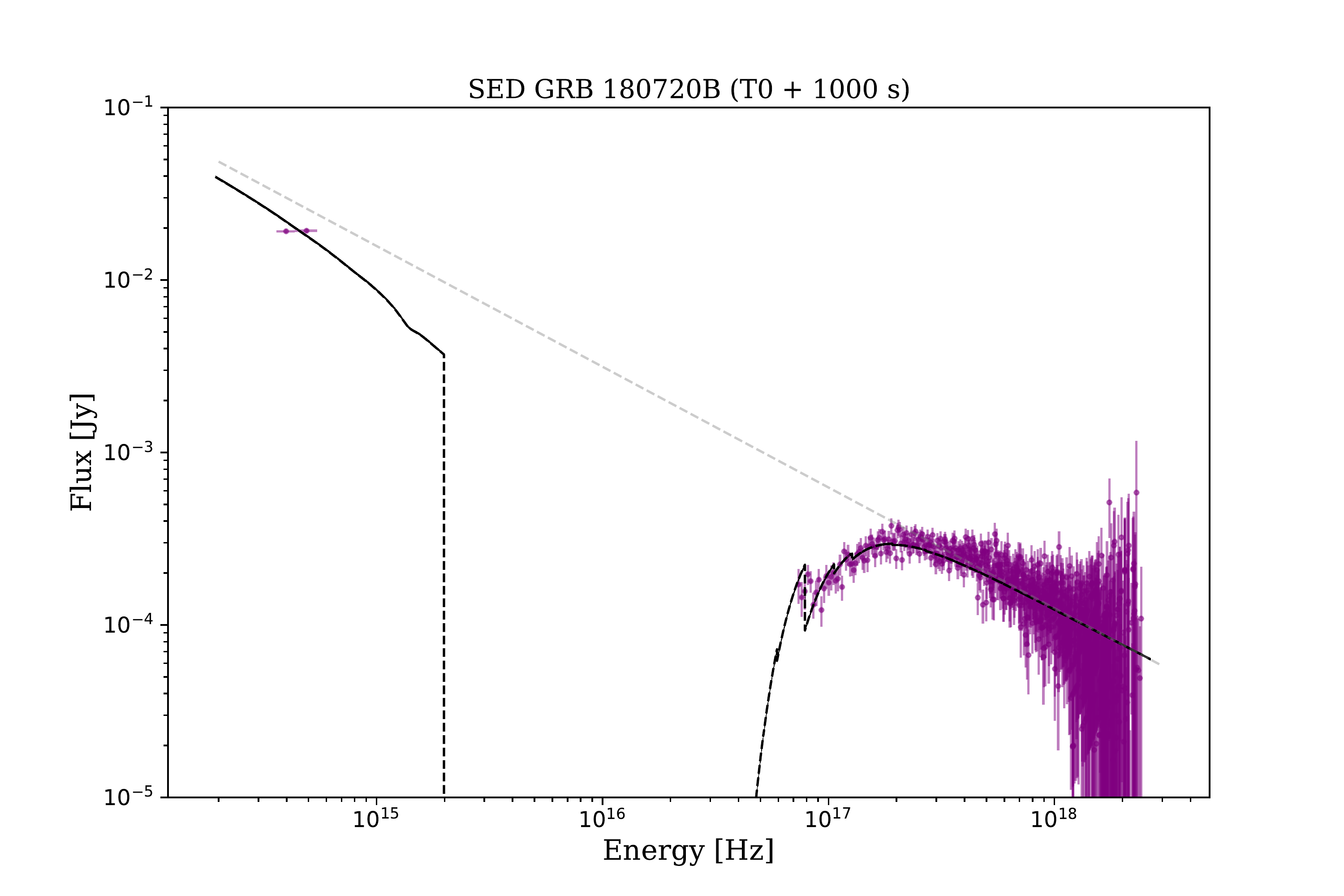}}
\resizebox*{0.49\textwidth}{0.36\textheight}
{\includegraphics{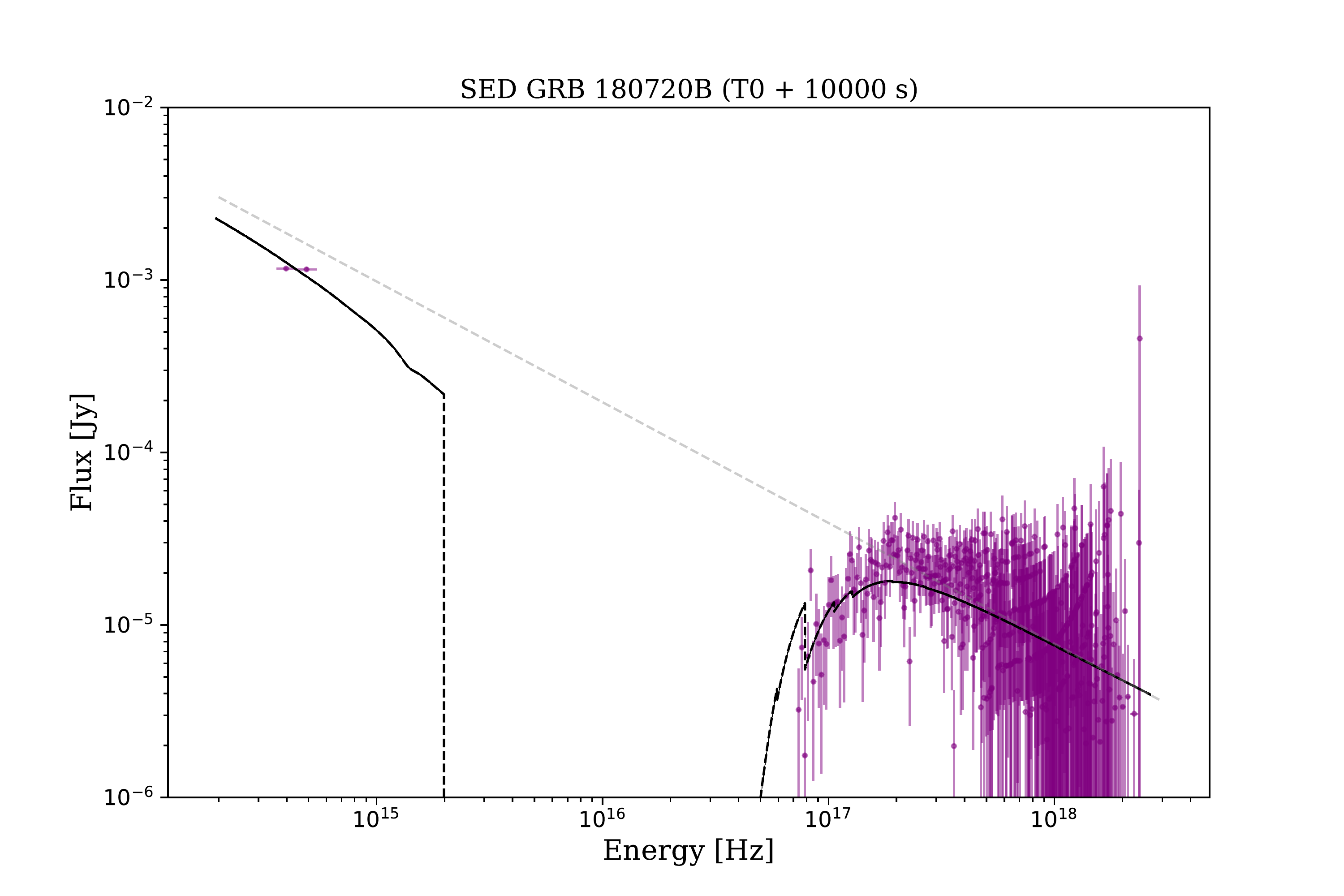}}\\
}
\caption{The broadband SEDs of the X-ray and optical observations are shown at 1000 s (left) and 10000 s (right) after the trigger time. The dashed gray lines are the best-fit curve from XSPEC.}
\label{sed}
\end{figure}

Figure \ref{sed} shows the broadband SEDs including the X-ray and optical observations at 1000 s (left) and 10000 s (right) with the best-fit PL with spectral indexes $0.68\pm0.06$ ($\chi^2/ndf=0.96$) and $0.70\pm0.05$ (0.97), respectively. The dashed gray lines correspond to the best-fit curves from XSPEC.\\
The left-hand panel in Figure \ref{grb180720B} shows the LAT, X-ray, optical and radio data with the best-fit PL functions given in Section 2. The LAT data is displayed at 100 MeV, X-ray at 1 keV, optical at the R-band and radio data at 15.5 and 1.4 GHz. The best-fit parameters of the temporal PL indexes obtained through the Chi-square $\chi^2$ minimization function are reported in Table \ref{table2}.   It is worth emphasizing that  radio data is not included in our analysis because there is only one data point for each energy band. \\ 
In order to analyze the LAT, X-ray and optical light curves we used the time intervals (epoch ``I", ``II", ``III" and ``IV") proposed for the X-ray light curve. Taking into account that to analyze epoch II, it is necessary to have the results of epochs III and IV, this epoch will be the last one to be analyzed.
\subsection{Epoch I:  $75 \,{\rm s}\lesssim t\lesssim 200\,{\rm s}$}
%
During this epoch,  the LAT and the optical light curves are modelled with PL functions and  the X-ray flare with two PLs. Considering that during this epoch the X-ray flare, the LAT and the optical light curves have different origins, we  first analyze the LAT and the optical light curves and then we examine the X-ray flare.   
\subsubsection{Analysis of LAT and optical light curves}
The best-fit values of the temporal and spectral PL indexes for the  LAT observations are $\alpha_{\rm LAT}=1.45\pm0.53$ and $\beta_{\rm LAT}=1.17\pm 0.15$, respectively, and the temporal index for the optical observations is $\alpha_{\rm O}=1.22\pm 0.02$. Taking into account that the LAT observations can be described by the third PL segment of the synchrotron forward-shock model,  and also that its temporal index is larger than the index of the optical observations ($\Delta \alpha \approx 0.3$), the optical observation can be described by the second PL segment  ($\propto t^{-\frac{(3p-3)}{4}}\epsilon^{-\frac{p-1}{2}}_\gamma$) of the synchrotron forward-shock model in the homogeneous medium.  In this case, the electron spectral index that explains both the LAT and optical observations would be  $p\approx2.6\pm 0.2$ and  $p\approx2.62\pm 0.02$, respectively, when the synchrotron emission radiates  in the homogeneous medium.  In the case of the afterglow wind model, the temporal index of the optical observations is usually larger than the one of the LAT observations.\\
\subsubsection{Analysis of the X-ray flare}
We used two PLs to fit the X-ray flare empirically  \citep[e.g. see,][]{2019ApJ...872..118B}.  Therefore, the X-ray flare is defined by the rise and decay  of the temporal indexes by -$(2.05\pm0.27$) and $2.74\pm0.06$, respectively, and a variability timescale of $\delta t/t\sim 1$. These values are discussed in terms of the reverse-shock emission and late central-engine activity. \\
\paragraph{Reverse-shock emission}
A reverse shock is believed to occur in the interaction between the expanding relativistic outflow and the external circumburst medium.  During this shock,  relativistic electrons heated and cooled down mainly by synchrotron and Compton scattering emission generate a single flare emission  \citep[see, e.g.,][]{2007ApJ...655..391K,  2012ApJ...751...33F, 2017arXiv171008514F}.  The evolution of reverse-shock emission is considered in the  thick and thin shell regimes, depending on the crossing time and the duration of the prompt phase \citep[e.g. see, ][]{2003ApJ...597..455K}. In the thick shell, the flare is  overlapped with the prompt emission and in the thin shell it is separated from the prompt phase.  Since the X-ray flare in GRB 180720B took place later than the burst emission,  the reverse-shock emission must evolve in the thin shell.\\
\cite{2007ApJ...655..391K} discussed the generation of an X-ray flare by Compton scattering emission in the early afterglow phase when the reverse shock is originated in the homogeneous medium and evolves in the thin shell. These authors found that the X-ray emission created in the reverse-shock region displays a time variability scale of $\delta t/t\sim 1$ and  varies as $F^{\rm ssc}_{\rm \nu, r}\propto t^{\frac{5(p-1)}{4}}$ before the peak and $\propto t^{-\frac{3p+1}{3}}$ after the peak. Taking into account the best-fit values  of the rise and decay indexes, the electron spectral indexes are $2.64\pm0.22$ and $2.42\pm 0.06$, respectively.\\
Considering the reverse shock evolving in a thin shell and in a homogenous medium, the Lorentz factor is bounded by the critical Lorentz factor $\Gamma < \Gamma_{\rm c}$  \citep{2003ApJ...595..950Z} and the deceleration time $t_{\rm dec} \approx 130\,{\rm s}$. The critical Lorentz factor and deceleration time scale are defined by {\small $\Gamma_{\rm c}=\left(\frac{3}{32\pi m_p}\right)^\frac18\,(1+z)^{\frac38}\,E^{\frac18}\,n^{-\frac18}\,T^{\frac38}_{90}$} and {\small $t_{\rm dec}\approx \left(\frac{3}{32\pi m_p}\right)^\frac13\,(1+z)\,E^{\frac13}\, n^{-\frac13}\,\Gamma^{\frac83}$}, respectively, where $E$ is the equivalent kinetic energy obtained using  the isotropic energy  and the efficiency to convert the kinetic energy into photons, $m_p$ is the proton mass and $T_{90}$ is the duration of the burst.  The maximum flux can be calculated by {\small $F^{\rm ssc}_{\rm \nu, r} = F^{\rm ssc}_{\rm max, r} \left( \frac{\epsilon_{\rm \gamma}}{\epsilon^{\rm ssc}_{\rm c, r}}\right)^{-\frac12}$} with $F^{\rm ssc}_{\rm max, r}$  and $\epsilon^{\rm ssc}_{\rm c, r}$  the maximum flux and  the cutoff energy break of the SSC emission, respectively \citep{2003ApJ...595..950Z,2016ApJ...831...22F}. \\
\paragraph{Late central-engine activity.}  In the framework of late central-engine activity, the ultra-relativistic jet has  several  mini-shells and the X-ray flare is the result of multiple internal shell collisions. The light curve is built as the superposition of the prompt emission from the late activity and the standard afterglow. In this context, the fast rise is naturally explained in terms of the short time-variability of the central engine $\delta t/t\ll 1$.   For a random magnetic field caused by internal shell collisions, the flux decays as $F_{\rm \nu, is} \propto t^{-2 p}$ in the slow cooling regime \citep[see e.g.;][]{2006ApJ...642..354Z}. In this case, the electron spectral index would correspond to an atypical value of $p=1.37\pm 0.03$.\\  
Given the temporal analysis, we conclude that the X-ray flare is most consistently explained by the reverse-shock emission rather than by the late activity of the central engine.
\subsection{Epoch III: $2.5\times 10^3 \,{\rm s} \lesssim t\lesssim 2.6\times 10^5\,{\rm s}$}
During this epoch, the spectral analysis indicates that the optical and X-ray observations are described with a PL with index $\beta_{\rm X,III}=0.70\pm 0.05$. The temporal analysis shows that the indexes of optical ($\alpha_{\rm O}=1.22\pm0.02$) and X-ray ($\alpha_{X, III}=1.26\pm0.06$) observations are consistent.    Therefore, the optical and X-ray fluxes evolve in the second PL segment of synchrotron emission in the homogeneous medium  for predicted values of $p\approx 2.62\pm 0.02$ and $p\approx2.68\pm 0.08$, respectively.  In the context of the X-ray light curve, this phase is known as the normal decay \citep[e.g. see][]{2006ApJ...642..354Z}. 
\subsection{Epoch IV: $ t\gtrsim 2.6\times 10^5\,{\rm s}$}
 The X-ray light curve during this time interval decays with $\alpha_{\rm X, IV}=1.70\pm 0.19$, which is consistent with the LAT light curve reported in epoch II. Taking into account epoch III, the temporal PL index varied as $\Delta \alpha \approx 0.45$ (from $\alpha_{\rm X,II}=1.26\pm0.06$ to $1.70\pm0.19$), which is consistent with the evolution  from the second to the third PL segments of synchrotron emission in the homogeneous medium for $p\approx 2.60\pm0.2$. Therefore,  the break observed in the X-ray light curve  during the transition from epochs III to IV can be explained as the transition of the synchrotron energy break below the X-ray observations at 1 keV. \\
\subsection{Epoch II: $200\, {\rm s} \lesssim t\lesssim 2500\,{\rm s}$}
In order to describe the LAT, X-ray and optical light curves correctly during this time interval, epochs I and III are taken into account.   Temporal and spectral analysis of the X-ray light curve  shows that during epoch II the PL indexes are $\alpha_{\rm X, II}=0.79\pm 0.08$ and $\beta_{\rm X, II}=0.68\pm 0.06$, respectively.      The spectral indexes associated with the X-ray observations during epochs II and III are very similar ( $\beta_{\rm X, II} \approx \beta_{\rm X, III}$),  and the spectral and temporal PL indexes of LAT and optical observations during epochs I and II are unchanged.    Taking into account that $\beta_{\rm X, II} \approx \beta_{\rm X, III}$, that there are no breaks in the LAT and optical light curves  and that the value of the temporal decay index is followed by the normal decay phase in the X-ray light curve,  epoch II is consistent with  the shallow  ``plateau" phase \citep[e.g. see][]{2009grb..book.....V}.  It is worth highlighting  that during this transition there was no variation of the spectral index.   A priori, we could think that X-ray observations during this epoch could be associated with the  second PL segment ($\propto t^{-\frac{(3p-3)}{4}}\epsilon^{-\frac{p-1}{2}}_\gamma$) of synchrotron emission. In this case the spectral index of the electron population, taking into account the temporal and spectral analysis, would be $p=2.05\pm0.11$ and $p=0.53\pm0.11$, respectively, which is different from the LAT and optical observations derived in the previous subsection. Hence, this hypothesis is rejected and we postulate the  ``plateau" phase.\\
The temporal and spectral theoretical indexes obtained by the evolution of the standard synchrotron model in  the  homogeneous medium are reported in Table \ref{table2}.  Theoretical and observational spectral and temporal indexes are consistent for $p\approx 2.6\pm0.2$. \\ 
\section{Discussion and description of Radio wavelengths and VHE gamma-rays}
\begin{figure}
	{ \centering
		\resizebox*{\textwidth}{0.7\textheight}
		{\includegraphics[angle=-90]{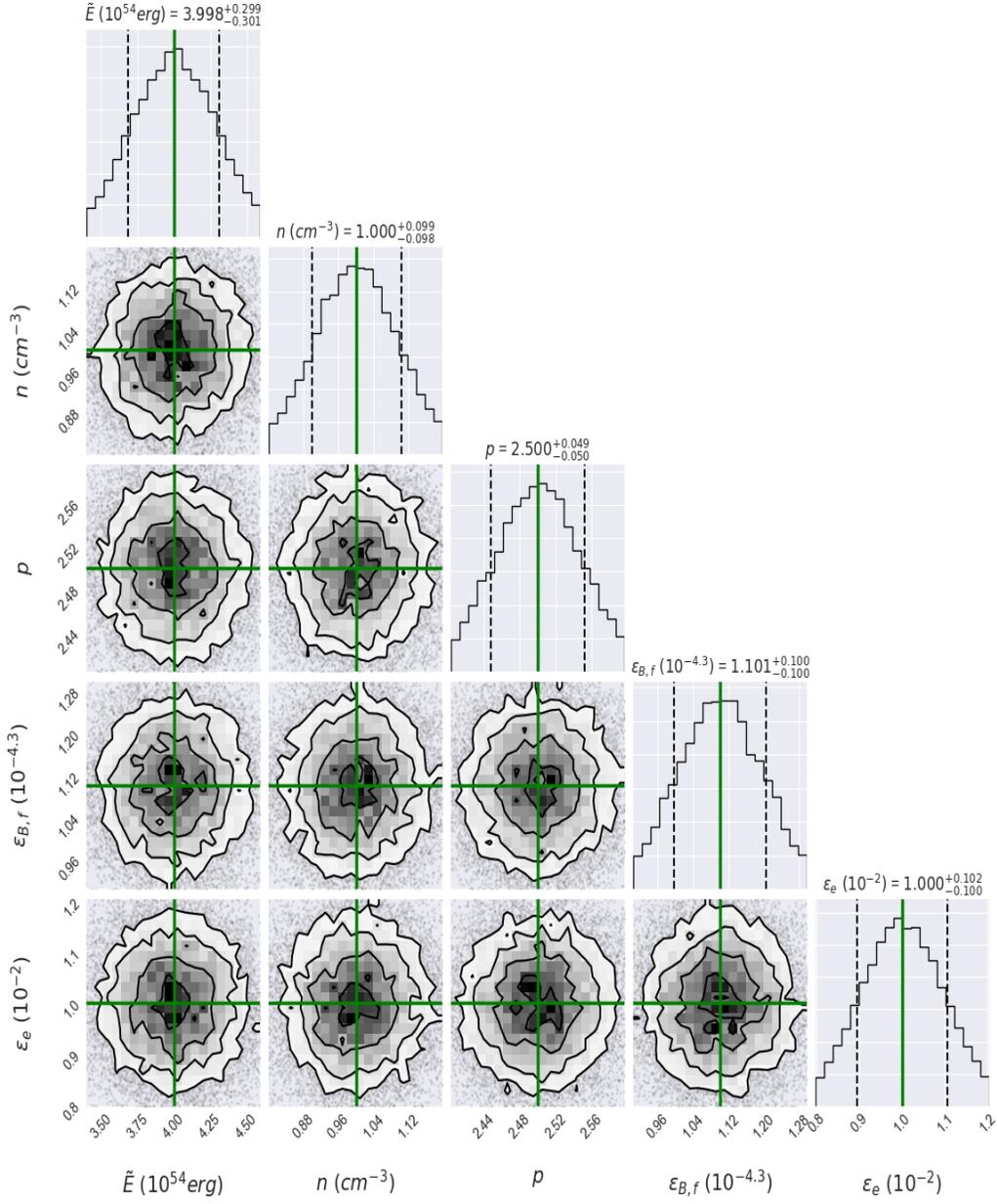}}
	}
	\caption{Best fits results for the LAT light curve at 100 MeV using our model and the MCMC calculations for GRB 180720B. The "corner plots" exhibit  the results obtained from the MCMC simulation. Labels above the 1-D KDE plot illustrate the 15\%, 50\% and 85\% quantiles for all parameters. The best-fit values are shown in green and reported in  column 2 of Table \ref{table3}.}
	\label{fig:param_LAT}
\end{figure}

\begin{figure}
	{ \centering
		\resizebox*{\textwidth}{0.7\textheight}
		{\includegraphics[angle=-90]{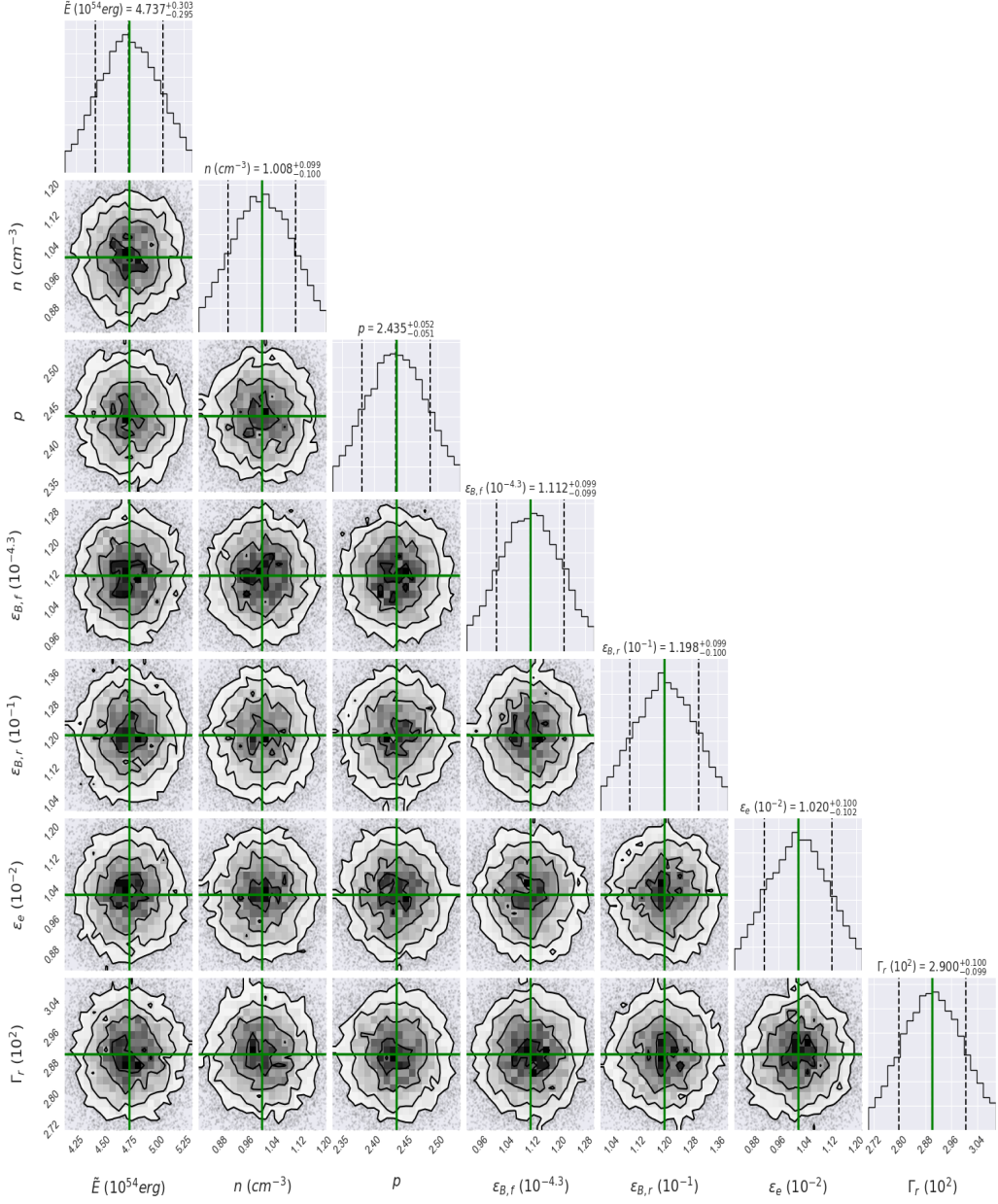}}
	}
	\caption{Same as Figure \ref{fig:param_LAT} but for the X-ray light curve at 10 keV. The best-fit values are shown in green and reported in column 3 of Table \ref{table3}.}
	\label{fig:param_xray}
\end{figure}

\begin{figure}
	{ \centering
		\resizebox*{\textwidth}{0.7\textheight}
		{\includegraphics[angle=-90]{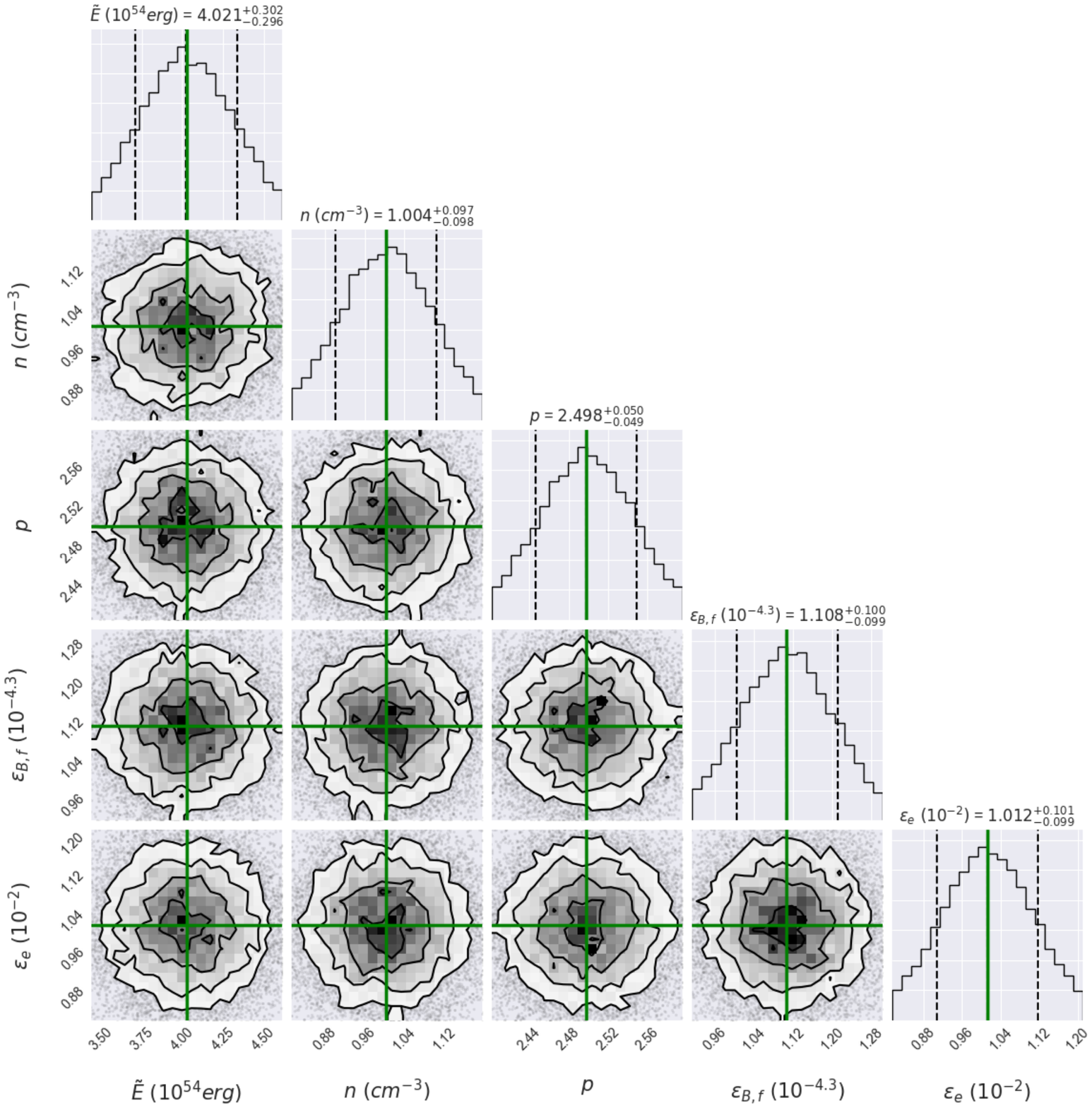}}
	}
	\caption{Same as Figure \ref{fig:param_LAT} but for the optical  light curve at 1 eV. The best-fit values are shown in green and reported in column 4 of Table \ref{table3}.}
	\label{fig:param_optical}
\end{figure}

We have shown that the temporal and spectral analysis of the multiwavelength (LAT, X-rays and optical bands) afterglow observed in GRB 180720B is consistent with the closure relations of the synchrotron forward-shock model evolving in a homogeneous medium.  Additionally, we have shown that the X-ray flare is consistent with the SSC reverse-shock model evolving in the thin shell in a  homogeneous medium.      In order to describe the LAT, X-ray and optical observations with our model,  we have constrained the electron spectral index, the microphysical parameters and the circumburst density using the Bayesian statistical technique  based on the Markov chain Monte Carlo (MCMC) method \cite[see][]{2019ApJ...871..200F}. These parameters were found by normalizing the PL segments at $\epsilon_\gamma=$ 100 MeV, 1 keV and 1eV  for  LAT, X-ray and optical observations, respectively.     We have used the synchrotron and SSC light curves in the slow-cooling regime when the outflow is decelerated in a homogeneous medium and the reverse shock evolves in thin shell.  The values reported for GRB 180720B such as the redshift $z=0.65$ \citep{2018GCN.22996....1V}, the equivalent isotropic energy $3\times 10^{53}\,{\rm erg}$ and the duration of the prompt emission $T_{90}=50\,{\rm s}$  \citep{2018GCN.22981....1R}  were used.  In order to compute the luminosity distance, the values of the cosmological parameters derived by \cite{2018arXiv180706209P} were used (Hubble constant $H_0=(67.4\pm 0.5)\,{\rm km\,s^{-1}\,Mpc^{-1}}$ and the matter density parameter $\Omega_{\rm m}=0.315\pm 0.007$).   The equivalent kinetic energy was obtained using  the isotropic energy  and the efficiency to convert the kinetic energy  into photons \citep{2015MNRAS.454.1073B}. \\
The best-fit value of each parameter found with our MCMC code is shown with a green line in Figures \ref{fig:param_LAT}, \ref{fig:param_xray} and \ref{fig:param_optical}  for LAT, X-ray and optical observations, respectively.  A total of 16000 samples with 4000 tuning steps were run.   The best-fit values for GRB 180720B  are reported in Table \ref{table3}.  The obtained values are  similar to those reported by other powerful GRBs  \citep{2010ApJ...716.1178A, 2013ApJ...763...71A, 2014Sci...343...42A, 2015ApJ...804..105F, 2016ApJ...818..190F, 2016ApJ...831...22F, 2017ApJ...848...94F, 2019arXiv190406976F}.  Given the values of the observed quantities and the best-fit values reported in Table \ref{table3}, the results are discussed as follows.\\
\\
\subsection{Describing the Radio emission}
During the afterglow, the self-absorption energy break lies in the radio bands and falls into two groups depending on the regime (fast or slow cooling) of the electron energy distribution.  For ${\rm min} \{\epsilon^{\rm syn}_{\rm a, f}, \epsilon^{\rm syn}_{\rm m, f}\}<\epsilon^{\rm syn}_{\rm c, f}$,  the observed synchrotron radiation flux can be in the weak ($\epsilon^{\rm syn}_{\rm a, f} \leq\epsilon_\gamma\leq\epsilon^{\rm syn}_{\rm m, f}$) or strong ($\epsilon^{\rm syn}_{\rm m, f} \leq\epsilon_\gamma\leq\epsilon^{\rm syn}_{\rm a, f}$) absorption regime \citep[e.g. see,][]{2013NewAR..57..141G}.
\paragraph{Weak absorption regime} In this case the synchrotron spectrum in the radio bands can be written as
{\small
\begin{eqnarray}
\label{radio_s1}
F^{\rm syn}_{\rm \nu, f}= F^{\rm syn}_{\rm \nu, max}  \cases{ 
\,\left(\frac{\epsilon^{\rm syn}_{\rm a, f}}{\epsilon^{\rm syn}_{\rm m, f}}\right)^\frac13 \,\left(\frac{\epsilon_\gamma}{\epsilon^{\rm syn}_{\rm a}}\right)^2,\hspace{0.3cm} \epsilon_\gamma<\epsilon^{\rm syn}_{\rm a,f} , \cr
\,\left(\frac{\epsilon_\gamma}{\epsilon^{\rm syn}_{\rm m, f}}\right)^\frac13 \,\, \hspace{1.5cm}       \epsilon^{\rm syn}_{\rm a,f}   <\epsilon_\gamma<\epsilon^{\rm syn}_{\rm m,f}\,,\hspace{.3cm} \cr
}
\end{eqnarray}
}
where the self-absorption energy break is
{\small
\be\label{Ea1}
 \epsilon^{\rm syn}_{\rm a, f}\propto (1+z)^{-\frac85} \epsilon^{-1}_{\rm e}\,\epsilon^\frac15_{\rm B, f} \Gamma^\frac85\,n^\frac45\,t^\frac35\,.
 \ee
 }
From eqs. \ref{radio_s1} and \ref{Ea1},  the radio light curve becomes $F^{\rm syn}_{\rm \nu, f}\propto:$  $t^\frac12 \epsilon^2_\gamma$ for $\epsilon_\gamma < \epsilon^{\rm syn}_{\rm a, f}$ and  $t^\frac12 \epsilon^\frac13_\gamma$ for $\epsilon^{\rm syn}_{\rm a, f}< \epsilon_\gamma < \epsilon^{\rm syn}_{\rm m, f}$.\footnote{We have used $\epsilon^{\rm syn}_{\rm m,f}\propto t^{-\frac32}$ and $F_{\rm \nu, max}\propto t^0$ with the bulk Lorentz factor  $\Gamma\propto t^{-\frac38}$ \citep{1998ApJ...497L..17S}} 
\paragraph{Strong absorption regime} In this case, the synchrotron spectrum in the radio bands is 
{\small
\begin{eqnarray}
\label{radio_s2}
F^{\rm syn}_{\rm \nu, f}= F^{\rm syn}_{\rm \nu, max}  \cases{ 
 \left(\frac{\epsilon^{\rm syn}_{\rm m, f}}{\epsilon^{\rm syn}_{\rm a, f}} \right)^{\frac{p+4}{2}}  \,\left(\frac{\epsilon_\gamma}{\epsilon^{\rm syn}_{\rm m, f}}\right)^2,\hspace{0.45cm} \epsilon_\gamma<\epsilon^{\rm syn}_{\rm m,f} , \cr
 \left(\frac{\epsilon^{\rm syn}_{\rm a, f}}{\epsilon^{\rm syn}_{\rm m, f}} \right)^{-\frac{p-1}{2}}  \,\left(\frac{\epsilon_\gamma}{\epsilon^{\rm syn}_{\rm a, f}}\right)^\frac52 \,\, \hspace{0.3cm}       \epsilon^{\rm syn}_{\rm m,f}   <\epsilon_\gamma<\epsilon^{\rm syn}_{\rm a,f}\,,\hspace{.3cm} \cr
}
\end{eqnarray}
}
where the self-absorption energy break in this case is  
\be\label{Ea2}
\epsilon^{\rm syn}_{\rm a, f}\propto (1+z)^{-\frac{p+6}{p+4}} \epsilon^{-\frac{2(p-1)}{p+4}}_{\rm e}\,\epsilon^{\frac{p+2}{2(p+4)}}_{\rm B, f} \Gamma^{\frac{4(p+2)}{p+4}}\,n^{\frac{p+6}{2(p+4)}}\,t^{\frac{2}{p+4}}\,.
\ee
From eqs. \ref{radio_s2} and \ref{Ea2}, the radio light curve becomes  $F^{\rm syn}_{\rm \nu, f}\propto:$  $t^\frac12 \epsilon^2_\gamma$ for $\epsilon_\gamma < \epsilon^{\rm syn}_{\rm m, f}$ and $ t^\frac54 \epsilon^\frac52_\gamma$ for $\epsilon^{\rm syn}_{\rm m, f}< \epsilon_\gamma < \epsilon^{\rm syn}_{\rm a, f}$.\\
%
\\
Taking into account the best-fit values reported in Table \ref{table3}, the synchrotron energy breaks  are $\epsilon^{\rm syn}_{\rm m, f}=25.43\,{\rm GHz}$, $\epsilon^{\rm syn}_{\rm a, f}=13.6\,{\rm GHz}$ (weak absorption) and $\epsilon^{\rm syn}_{\rm a, f}=2.4\times10^3\,{\rm GHz}$ (strong absorption) at 1 day, and $\epsilon^{\rm syn}_{\rm m, f}=0.8\,{\rm GHz}$, $\epsilon^{\rm syn}_{\rm a, f}=13.6\,{\rm GHz}$  (weak absorption)  and $\epsilon^{\rm syn}_{\rm a, f}=4.5\times10^2\,{\rm GHz}$ (strong absorption) at 10 days.  Clearly, the synchrotron spectrum lies in the regime $\epsilon^{\rm syn}_{\rm m,f} \leq\epsilon_\gamma\leq\epsilon^{\rm syn}_{\rm a, f}$ and radio observations are in the second PL segment.\\
Using the best-fit parameters obtained with our MCMC code, we describe the radio data points as shown in the right-hand panel of Figure \ref{grb180720B}.  In order to describe the radio data at 15.5 and 1.4 GHz with a PL, we multiply the radio data point at 1.4 GHz by 25 and we also normalize the synchrotron flux at the same radio band.
\subsection{Describing the LAT photons}
\begin{figure}[h!]
{ \centering
\resizebox*{0.7\textwidth}{0.35\textheight}
{\includegraphics{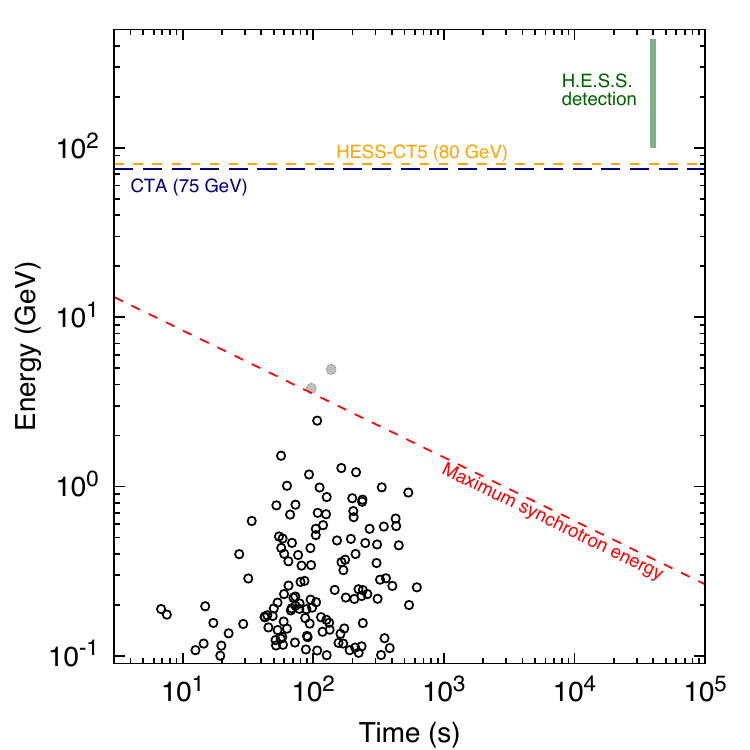}}
}
\caption{All the photons with energies $> 100$ MeV and probabilities $>90$\% of being associated with  GRB 180720B.  The red dashed line represents the maximum photon energies released by the synchrotron forward-shock model.  Photons with energies above the synchrotron limit  are in gray and below are in black. The sensitivities of CTA and HESS-CT5  observatories at 75 and 80 GeV are shown  in yellow and blue dashed lines, respectively \citep{2016CRPhy..17..617P}.}
\label{photons_MeV}
\end{figure}

\begin{figure}[h!]
{ \centering
\resizebox*{0.7\textwidth}{0.35\textheight}
{\includegraphics{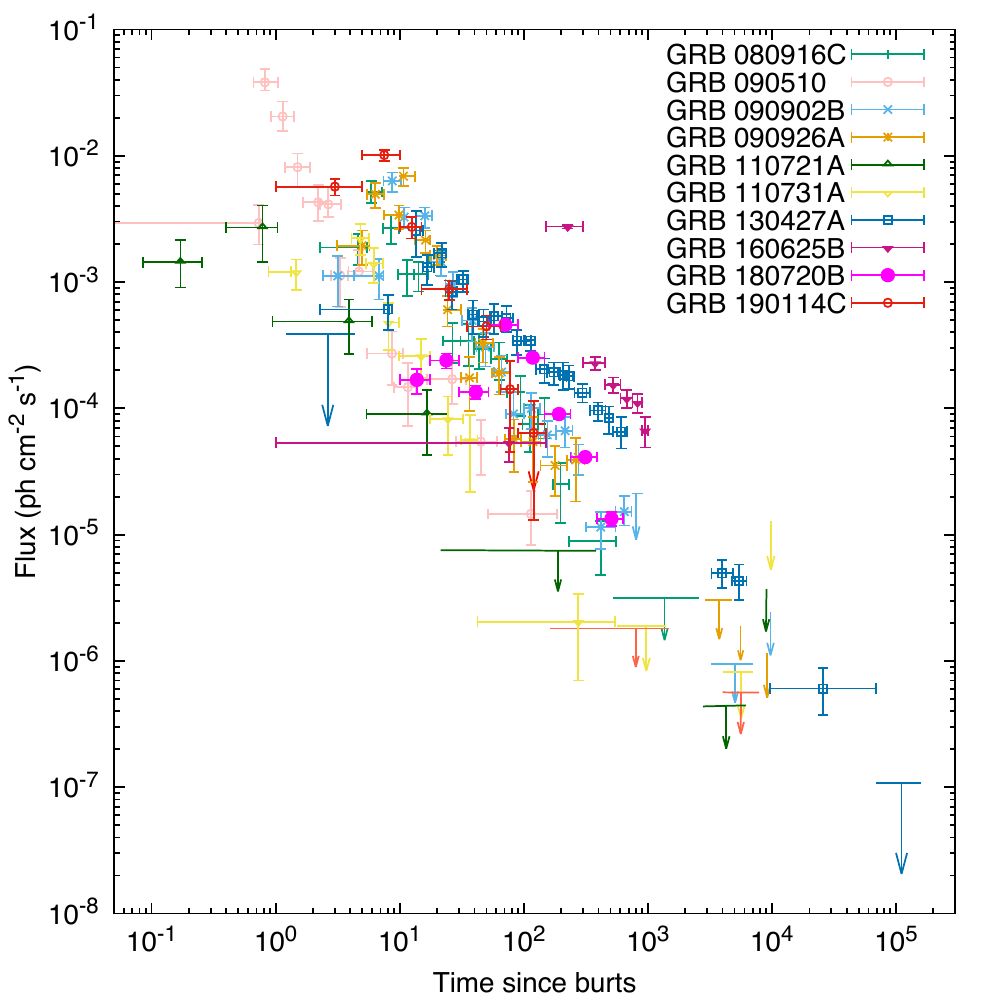}}
}
\caption{Comparison of the Fermi-LAT photon flux light curve from GRB 180720B (blue circles) with other LAT-detected bursts.  LAT data of energetic GRBs are taken from \cite{2013ApJS..209...11A,2013ApJ...763...71A,  2014Sci...343...42A, 2019arXiv190406976F}.}
\label{all_GRBs}
\end{figure}

Using the best-fit value of the homogeneous density found with our MCMC code we plot, in Figure \ref{photons_MeV},  the evolution of the maximum photon energy radiated by synchrotron emission from the forward-shock region (red dashed line) and  all individual photons with probabilities $>$ 90\% to be associated to GRB180720B and  energies above $\gtrsim 100$ MeV.   Photons with energies above the synchrotron limit  are in gray and the ones with energy below this limit are in black.  The sensitivities of CTA and HESS-CT5  observatories at 75 and 80 GeV are shown  in yellow and blue dashed lines, respectively \citep{2016CRPhy..17..617P}. The emission reported by the HESS Collaboration during the CTA symposium \footnote{https://indico.cta-observatory.org/event/1946/timetable/} is shown in green. Figure \ref{photons_MeV} shows that all photons cannot be interpreted in the standard synchrotron forward-shock model.   Although this burst is a good candidate for accelerating particles up to very-high energies and then producing TeV neutrinos, no neutrinos were spatially or temporally associated with this event. This negative result could be explained in terms of the low amount of baryon load  in the outflow. In this case the production of VHE photons favors  leptonic over hadronic models.   Therefore, we propose that LAT photons above the synchrotron limit would be interpreted in the SSC framework.    It is worth highlighting that the LAT photons below the  synchrotron limit (the red dashed line) could be explained in the standard synchrotron forward-shock model and beyond this  limit the SSC model would describe the LAT photons. For instance, a superposition of synchrotron and SSC emission originated in the forward-shock region could be invoked to interpret the LAT photons \citep[e.g., see][]{2015MNRAS.454.1073B}.\\
\\
The Fermi-LAT photon flux light curve of GRB 180720B presents characteristics  similar to other LAT-detected GRBs, such as GRB 080916C \citep{2009Sci...323.1688A}, GRB 090510 \citep{2010ApJ...716.1178A}, GRB 090902B \citep{2009ApJ...706L.138A}, GRB 090926A \citep{2011ApJ...729..114A} GRB 110721A \citep{2013ApJ...763...71A}, GRB 110731A \citep{2013ApJ...763...71A}, GRB 130427A \citep{2014Sci...343...42A}, GRB 160625B \citep{2017ApJ...848...15F} and GRB 190114C \citep{2019arXiv190406976F}, as shown in Figure \ref{all_GRBs}.  All of these GRBs exhibited VHE photons and a long-lived emission lasting more than the prompt phase. This figure shows that during the prompt phase, the high-energy emission from GRB 180720B is one of the weakest and during the afterglow it is one of the strongest.\\
\subsection{Production of VHE gamma-rays to be detected by GeV - TeV observatories}
\begin{figure}[h!]
{\centering
\resizebox*{0.45\textwidth}{0.33\textheight}
{\includegraphics{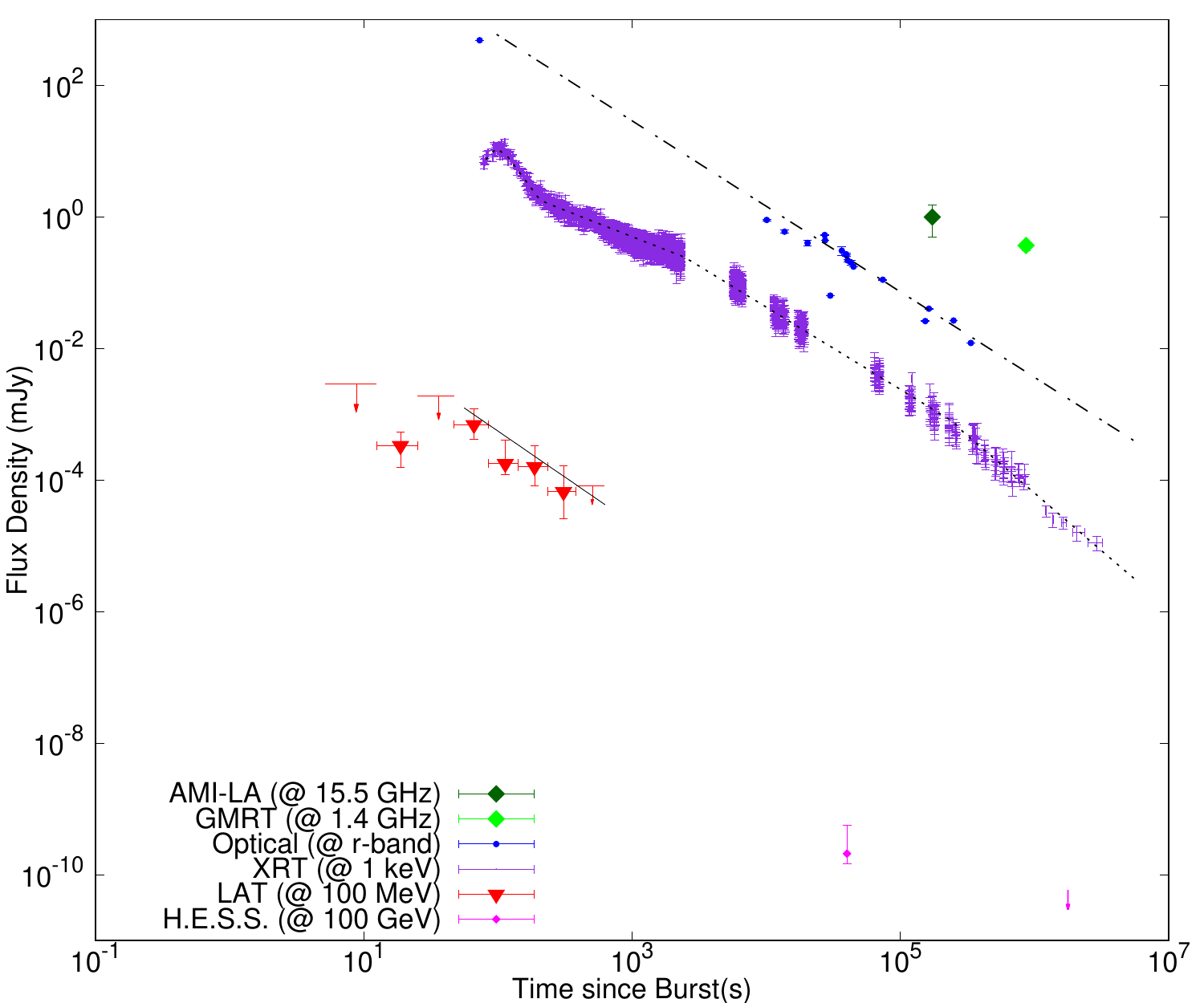}}
\resizebox*{0.45\textwidth}{0.33\textheight}
{\includegraphics{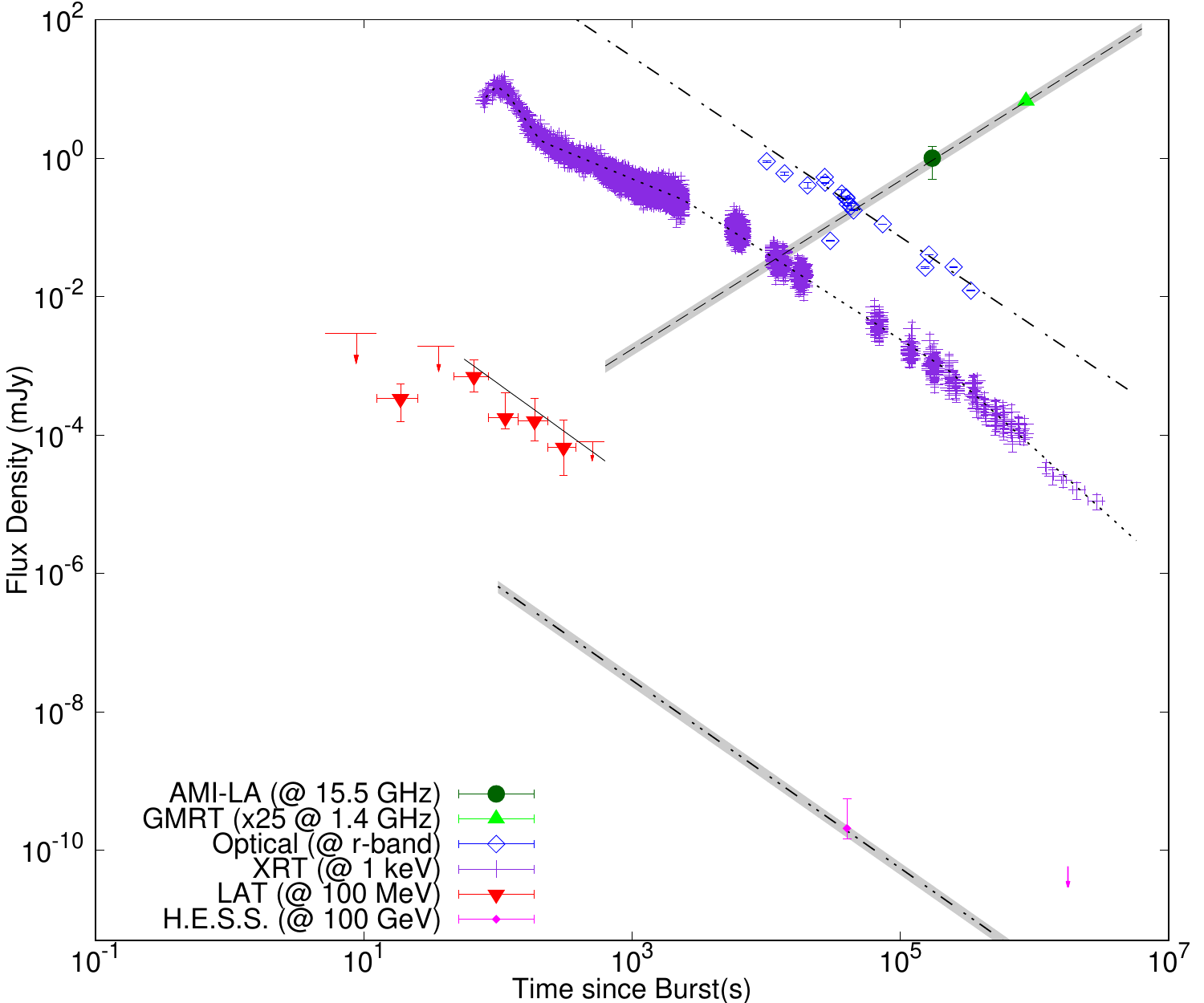}}\\
}
\caption{The left-hand panel shows multiwavelength light curves and fits of the LAT, X-ray and optical observations of GRB 180720B.  The right-hand panel shows  the same as the left-hand panel but with the description of radio wavelengths and the VHE gamma-rays at 100 GeV.}
\label{grb180720B}
\end{figure}

The dynamics of the synchrotron forward-shock emission in a homogeneous medium have been widely explored \citep[e.g. see,][]{1998ApJ...497L..17S}.   Synchrotron photons radiated in the forward shocks can be up-scattered by the same electron population.  The inverse Compton scattering model has been described in \cite{2000ApJ...544L..17P, 2000ApJ...532..286K}.   Given the energy breaks, the maximum flux, the spectra and the light curves of the synchrotron radiation, the SSC light-curves for the fast- and slow-cooling regime become 
{\small
\begin{eqnarray}
\label{fcssc_t}
F^{\rm ssc}_{\rm \nu, f}\propto \cases{ 
\,t^{\frac{1}{3}}\, \epsilon_\gamma^{\frac13},\hspace{1.3cm} \epsilon_\gamma<\epsilon^{\rm ssc}_{\rm c, f}, \cr
t^{\frac{1}{8}}\, \epsilon_\gamma^{-\frac12},\hspace{1.1cm} \epsilon^{\rm ssc}_{\rm c, f}<\epsilon_\gamma<\epsilon^{\rm ssc}_{\rm m, f}, \cr
t^{-\frac{9p-10}{8}}\,\epsilon_\gamma^{-\frac{p}{2}},\,\,\,\,\,\,\epsilon^{\rm ssc}_{\rm m, f}<\epsilon_\gamma\,, \cr
}
\end{eqnarray}
}
and
{\small
\begin{eqnarray}
\label{scssc_t}
F^{\rm ssc}_{\rm \nu, f}\propto \cases{
t\,\epsilon_\gamma^{\frac13},\hspace{2.1cm} \epsilon_\gamma<\epsilon^{\rm ssc}_{\rm m, f},\cr
t^{-\frac{9p-11}{8}}\epsilon_\gamma^{-\frac{p-1}{2}}, \hspace{0.7cm}  \epsilon^{\rm ssc}_{\rm m, f}<\epsilon_\gamma<\epsilon^{ssc}_{\rm c, f},\,\,\,\,\,\cr
t^{-\frac{9p-10}{8} + \frac{p-2}{4-p}}\,\epsilon_\gamma^{-\frac{p}{2}},\,\,\,\,\epsilon^{\rm ssc}_{\rm c, f}<\epsilon_\gamma\,, \cr
}
\end{eqnarray}
}
respectively, where the SSC energy breaks are 
{\small
\bary\label{sscf}
\epsilon^{\rm ssc}_{\rm m, f}&\propto&  (1+z)^{\frac54}\,\epsilon_{\rm e}^{4}\,\epsilon_{\rm B, f}^{\frac12}\,n^{-\frac14}\,E^{\frac34}\,t^{-\frac94}\cr
\epsilon^{\rm ssc}_{\rm c, f}&\propto& \, (1+z)^{-\frac34}\,(1+Y)^{-4}\,\epsilon_{\rm B, f}^{-\frac72}\,n^{-\frac94}\,E^{-\frac54}\,t^{-\frac14}\,,
\eary
}
with $Y$ the Compton parameter.  In the Klein-Nishina regime the break energy is given by
{
\small
\be
E^{\rm KN}_{\rm c}\propto (1+z)^{-1}\, (1+Y)^{-1} \,\epsilon_{\rm B,f}^{-1}\,n^{-\frac23}\,\Gamma^\frac23\,E^{-\frac13}\,t^{-\frac14}.
\ee
}
In this case the maximum flux emitted in this process is  $F^{\rm ssc}_{\rm \nu,max}\propto \,(1+z)^{\frac34}\,\epsilon_{\rm B, f}^{\frac12}\,n^{\frac54}\,d^{-2}\,E^{\frac54}\, t^{\frac14}$ where $d$ is the luminosity distance.\\
Given the best-fit parameters reported in Table \ref{table3}, the SSC light curve at 100 GeV is plotted in Figure \ref{grb180720B} (right).  The break energy in the KN regime is $ 851.7$ and  $478.9\,{\rm GeV}$ at 1 and 10 hours, respectively, which is above the flux at 100 GeV. The characteristic and cutoff break energies are $4.4 \times 10^3$ and $24. 7\,{\rm eV}$,  and $50.1$  and $43.2\,{\rm GeV}$ at 1 and 10 hours, respectively, which indicates that at 100 GeV, the SSC emission is evolving in the third PL segment of the slow-cooling regime.   The right-hand panel in Figure \ref{grb180720B} shows the SSC flux computed at 100 GeV with the parameters derived in our model using  the multiwavelength observations of GRB 180720B. Our model explains the VHE emission announced by HESS team during the high energy emission at the CTA symposium.\footnote{https://indico.cta-observatory.org/event/1946/timetable/}  It is worth nothing that the SSC flux equations are degenerate in the parameter values such that for a distinct set of parameter values similar results could be obtained, as shown in \cite{2019arXiv190511312W}.\\
\subsection{The magnetic microphysical parameters}
The best-fit parameters of the magnetic fields  found in the forward- and reverse-shock regions are different. The parameter associated with the magnetic field in the reverse shock lies in the range of the expected values for the reverse shock to be formed and leads to an estimate of the magnetization parameter of $\sigma\simeq 0.4$.  In the opposite situation (e. g. $\sigma\gg$1),  particle acceleration would hardly be  efficient and the X-ray flare  from the reverse shock would have been suppressed \citep{2004A&A...424..477F}.  Considering the microphysical parameter associated with the reverse-shock region, we found that the strength of the magnetic field in this region is stronger than the magnetic field in the forward-shock region ($\simeq 45$ times).  This suggests that the jet composition of GRB 180720B could be Poynting  dominated.  \cite{2005ApJ...628..315Z} described the emission generated in the reverse shock from an outflow with an arbitrary value of the magnetization parameter. They found that the Poynting energy is transferred to the medium only until the reverse shock has disappeared. Given the timescale of the reverse shock associated with the X-ray flare,  the shallow decay segment observed in the X-ray light curve of GRB 180720B  could be interpreted as the late transfer of the Poynting energy to the homogeneous medium.   These results agree with  some authors who claim that Poynting flux-dominated models with  a moderate degree of magnetization can explain the LAT observations in powerful GRBs \citep{2014NatPh..10..351U,2011ApJ...726...90Z}, and in particular the properties exhibited in the light curve of GRB 180720B.\\
\\
Using the synchrotron reverse-shock model \citep{2003ApJ...597..455K, 2000ApJ...545..807K} and the best-fit values (see Table \ref{table3}), the self-absorption, the characteristic and cutoff energy breaks of $3.4\times10^{-7}\,{\rm eV}$, $0.4\,{\rm eV}$ and $7.3\times 10^{-4}\,{\rm eV}$, respectively, indicate that the synchrotron radiation evolves in the fast-cooling regime.  Given that the self-absorption energy break is smaller than the cutoff and characteristic breaks, the synchrotron emission originated from the reverse-shock region lies in the weak self-absorption regime and hence, a thermal component in this region cannot be expected \citep{2003ApJ...597..455K}.\\
\\
The spectral and temporal analysis of the forward and reverse shocks at the beginning of the afterglow phase together with the best-fit value of the circumburst density  lead to an estimate of the initial bulk Lorentz factor, the critical Lorentz factor and the shock crossing time  $\Gamma\simeq$ 200, $\Gamma_{\rm c}\simeq330$ and $t_d\simeq 100\,{\rm s}$, respectively.  These values are consistent with the evolution of the reverse shock in the thin-shell case and the duration of the X-ray flare.\\
\\
\section{Conclusions}
GRB 180720B is a long burst detected and followed-up by a large number of observatories in multiwavelenths that range from radio bands to GeV gamma-rays.  The  simultaneous  GeV, gamma-ray, X-ray, optical  and radio bands are presented over multiple observational periods beginning just after the BAT trigger time and extending for more than 33 days.\\ %
The GBM light curve and spectrum were analyzed using the Band and Comptonized functions in the energy range of 10 - 1000 keV during the time interval [0.000, 60,416 s]. The light curve formed by a bright FRED-like peak and followed by two significant peaks is consistent with the prompt phase.  The Fermi LAT light curve and spectrum were derived around the reported position of GRB 180720B.   We have shown that the photon flux light curve recently reported in the second GRB catalog \citep{2019ApJ...878...52A} is consistent with the one obtained in this work.   The highest-energy photons with energies of 3.8 and 4.9 GeV detected by the LAT instrument at 97 and 138 s, respectively,  after the GBM trigger can hardly be interpreted in the standard synchrotron forward-shock model.    Photons below the synchrotron limit can be explained well by synchrotron emission from the forward shock.   The temporal and spectral indexes of the Fermi LAT observations are compatible and consistent with the synchrotron forward-shock afterglow. \\   
\\
The temporal and spectral analysis of the  X-ray observations suggested four different behaviours whereas the optical R-band observations just one.  We find that the X-ray flare is most consistently interpreted with the SSC model from the reverse shock region evolving in a thin shell. This model can explain the timescales, the maximum observed flux, and the rise and fall of temporal PL indexes.  The temporal decay index in the range between 0.2 and 0.8,  as found in a large fraction of bursts with no variation of the spectral index during the transition, is consistent with the shallow plateau phase \citep[e.g. see][]{2009grb..book.....V}. The temporal PL index  after the break is consistent with the normal decay  in a uniform IMS-like medium.   The chromatic break at $2\times 10^5$ s observed in the X-ray but not in optical light curve is consistent with the fact that the cooling energy break of the synchrotron model becomes less than the X-ray observations at 1 keV.\\
Temporal and spectral PL indexes observed in the LAT, X-rays and optical bands during different intervals favor the model of an afterglow in a homogeneous medium.  The best-fit parameters derived with our  MCMC code indicate that the outflow is endowed with magnetic fields, the radio data is in the self-absorption regime and the LAT photons above the synchrotron limit are consistent with SSC forward-shock model.   The SSC forward-shock model with our parameters can explain  the LAT photons beyond the synchrotron limit as well as the emission reported by the HESS Collaboration. The X-ray flare and the ``plateau" phase with their corresponding timescales could be explained by the late transfer of the magnetic energy  into the uniform medium, emphazising that the outflow is magnetized. \\
%

%
%
\acknowledgements
We thank Rodolfo Barniol Duran, Peter Veres, Alexander A. Kann, Michelle Hui,  Alan Watson, Fabio De Colle and Diego Lopez-Camara for useful discussions. NF  acknowledges  financial  support  from UNAM-DGAPA-PAPIIT  through  grant  IA102019.  BBZ acknowledges support from National Thousand Young Talents program of China and National Key Research and Development Program of China (2018YFA0404204) and The National Natural Science Foundation of China (Grant No. 11833003).

\clearpage
%
%
%
%

%
%
\newpage
\begin{table}
\centering \renewcommand{\arraystretch}{2}\addtolength{\tabcolsep}{3pt}
\caption{The best-fit values of the temporal PL indexes derived from the XRT light curve of GRB 180720B.}
\label{table1}
\begin{tabular}{c  c  c  c }
 \hline \hline
\scriptsize{X-rays} &\hspace{0.5cm}   \scriptsize{Interval}  &\hspace{0.5cm}   \scriptsize{Index}    & \hspace{0.5cm} \scriptsize{ $\chi^2$/ndf} \\ 
\scriptsize{} & \hspace{0.5cm} \scriptsize{(s)} & \hspace{0.5cm}  \scriptsize{($\alpha_{\rm X}$)}   &   \\ 
\hline \hline
\scriptsize{I}   	        & \hspace{0.5cm} \scriptsize{$\leq 1.3\times 10^2$}  &\hspace{0.5cm} \scriptsize{$-2.05\pm0.27$}		&\hspace{0.5cm}  \scriptsize{$1.12$}\\	
   		 	        & \hspace{0.5cm} \scriptsize{$> 1.3\times10^2$ }  &\hspace{0.5cm} \scriptsize{$2.74\pm0.06$}		&\hspace{0.5cm}  \scriptsize{$1.27$}\\	\cdashline{1-4}    
\scriptsize{II}   	        & \hspace{0.5cm} \scriptsize{$(0.2 - 2.5)\times 10^3$}  &\hspace{0.5cm} \scriptsize{$0.79\pm0.08$}		&\hspace{0.5cm}  \scriptsize{$1.31$}\\\cdashline{1-4}
\scriptsize{III}   	        & \hspace{0.5cm} \scriptsize{$(0.25 - 26.1) \times10^4$ }  &\hspace{0.5cm} \scriptsize{$1.26\pm0.06$}		&\hspace{0.5cm}  \scriptsize{$1.29$}\\\cdashline{1-4}
\scriptsize{IV}   	        & \hspace{0.5cm} \scriptsize{$\geq 2.6\times10^5$ }  &\hspace{0.5cm} \scriptsize{$1.75\pm0.09$}		&\hspace{0.5cm}  \scriptsize{$1.21$}\\
\hline \hline
\end{tabular}
\end{table}
%
%

%
\begin{table}
\centering \renewcommand{\arraystretch}{2}\addtolength{\tabcolsep}{-2pt}
\caption{The best-fit paramters of the spectral and temporal indexes using the LAT, X-ray and optical observations.  In addition, the theoretical predictions of  the spectral and temporal indexes are calculated for $p=2.6\pm0.2$.  }
\label{table2}
\begin{tabular}{ l r c r c r c r c}
\hline
\hline
		                            &   									  \normalsize{Observation} & \normalsize{Theory} &  \normalsize{Observation} & \normalsize{Theory}  &  \normalsize{Observation} & \normalsize{Theory} &\normalsize{Observation} & \normalsize{Theory}         \\
		         & 		I&    & II &  & III& &  IV &                                                                             \\
		                           
\hline \hline
\scriptsize{\bf{LAT flux}} 	& 			 				                         &	  		 	         &	   & 	& &		 \\ 
\hline \hline  
 \scriptsize{$\alpha_{\rm LAT}$}	 \hspace{0.1cm}	&		\scriptsize{$1.45\pm0.53$}  &  \scriptsize{$1.45\pm0.15$}     &		\scriptsize{$1.45\pm0.53$}	&		\scriptsize{$ 1.45\pm0.15 $ } & 	\scriptsize{$ - $ }&	\scriptsize{$ - $ }& 	\scriptsize{$ - $ }&	\scriptsize{$ - $ }\\
 \scriptsize{$\beta_{\rm LAT}$}	 \hspace{0.1cm}	&		\scriptsize{$1.17\pm0.15$}       &   \scriptsize{$1.30\pm0.10$}   & 	\scriptsize{$1.15\pm0.15$}& 		\scriptsize{$ 1.30\pm0.10 $} & 	\scriptsize{$ - $ }&	\scriptsize{$ - $ }  & 	\scriptsize{$ - $ }&	\scriptsize{$ - $ }	\\

\\
\hline \hline
\scriptsize{\bf{X-ray flux}}		         & 		&    &   & &                                                                               \\
\hline \hline
 \scriptsize{$\alpha_{\rm X}$ ($\lesssim 1.3\times10^2\,{\rm s}$)}	\hspace{0.1cm}		 &			\scriptsize{-$(2.05\pm 0.27)$} & \scriptsize{-$(2.00\pm0.25)$} 	& \scriptsize{$-$}  &  \scriptsize{$-$} &	\scriptsize{$-$ } & 	\scriptsize{$-$ }&	\scriptsize{$-$ } & 	\scriptsize{$-$ }\\
 \scriptsize{$\alpha_{\rm X}$ ($>1.3\times10^2\,{\rm s}$)}	\hspace{0.1cm}		 &			\scriptsize{$2.74\pm 0.08$} & \scriptsize{$3.03\pm0.25$} 	& \scriptsize{$0.79\pm0.08$}  &  \scriptsize{$\sim(0.5 - 0.8)$} &	\scriptsize{$1.26\pm0.06$ } & 	\scriptsize{$1.20\pm0.15$ }&	\scriptsize{$1.70\pm0.19$ } & 	\scriptsize{$1.45\pm0.15$ }\\

 \scriptsize{$\beta_{\rm X}$}       \hspace{0.1cm}	 &			\scriptsize{$-$} & \scriptsize{$-$}  & \scriptsize{$0.68\pm0.06$}  &   \scriptsize{$0.80\pm0.10$}	& 	\scriptsize{$0.70\pm0.05$} &	\scriptsize{$0.80\pm0.1-$ } &	\scriptsize{$-$ } & 	\scriptsize{$-$ }   \\
\\
\hline\hline
\scriptsize{\bf{Optical flux}}  		         &                                             &  &                                     \\
\hline\hline
  \scriptsize{$\alpha_{\rm O}$}   \hspace{0.1cm}	&	\scriptsize{$1.22\pm0.012$}  &\scriptsize{$1.20\pm0.15$}	& \scriptsize{$1.22\pm0.012$}  & \scriptsize{$1.20\pm0.15$} & 	\scriptsize{$1.22\pm0.012$}  & 	\scriptsize{$ 1.20\pm0.15 $ }& 	\scriptsize{$-$}  &	\scriptsize{$ - $ }\\
\scriptsize{$\beta_{\rm O}$}        \hspace{0.1cm}	&	\scriptsize{$-$}  & \scriptsize{$-$}         & \scriptsize{$0.68\pm 0.06$} &  \scriptsize{$0.80\pm0.10$} &	\scriptsize{$0.70\pm0.05$} &	\scriptsize{$ 0.80\pm0.10 $ } & 	\scriptsize{$ - $ }&	\scriptsize{$ - $ }	\\
\hline \hline
\end{tabular}
\end{table}

\newpage
\begin{table}
\centering \renewcommand{\arraystretch}{2}\addtolength{\tabcolsep}{3pt}
\caption{Median values of parameters found with symmetrical quantiles (15\%, 50\%, 85\%).\\ The external shock model was used to constrain the values of the parameters.}
\label{table3}
\begin{tabular}{ l  c c c  c c}
\hline
\hline
{\large   Parameters}	& 	&	{\large  Median}  & 		 				                           		 	        	   		 \\ 
                          	 	&  {\normalsize LAT (100 MeV)} & 	{\normalsize X-ray (1 keV)} & {\normalsize Optical (1 eV)} 	                           		 	        	   		 \\ 

\hline \hline
\\
\small{$E\,(10^{54}\,{\rm erg})$}	\hspace{1cm}&     \small{$3.998^{+0.299}_{-0.301}$}	\hspace{0.7cm} &  \small{$4.737^{+0.303}_{-0.100}$} \hspace{0.7cm}&  \small{$4.021^{+0.302}_{-0.296}$} 	 \\
\small{$n\,({\rm cm^{-3}})$}	\hspace{1cm}&     \small{$1.000^{+0.099}_{-0.098}$}	\hspace{0.7cm} &  \small{$1.008^{+0.099}_{-0.100}$} \hspace{0.7cm}&  \small{$1.004^{+0.097}_{-0.098}$}  \\
\small{$\epsilon_{\rm B,f}\,(10^{-4.3})$}	\hspace{1cm}&     \small{$1.101^{+0.100}_{-0.100}$}	\hspace{0.7cm} &  \small{$1.112^{+0.099}_{-0.099}$} \hspace{0.7cm}&  \small{$1.108^{+0.100}_{-0.099}$} 	 \\
\small{$\epsilon_{\rm e}\,(10^{-2})$}	\hspace{1cm}&     \small{$1.000^{+0.102}_{-0.100}$}	\hspace{0.7cm} &  \small{$1.020^{+0.100}_{-0.102}$} \hspace{0.7cm}&  \small{$1.012^{+0.101}_{-0.099}$} 	 \\
\small{$\epsilon_{\rm B,r}\,(10^{-1})$}	\hspace{1cm}&     \small{$-$}	\hspace{0.7cm} &  \small{$1.200^{+0.099}_{-0.098}$} \hspace{0.7cm}&  \small{$-$} 	 \\
\small{$p$}	\hspace{1cm}&     \small{$2.500^{+0.049}_{-0.050}$}	\hspace{0.7cm} &  \small{$2.435^{+0.052}_{-0.051}$} \hspace{0.7cm}&  \small{$2.498^{+0.050}_{-0.049}$}  \\
\small{$\Gamma_r\, (10^2)$ } \hspace{1cm} &     \small{$-$}	\hspace{0.7cm}  &  \small{$2.900^{+0.100}_{-0.099}$} \hspace{0.7cm}&  \small{$-$}  \\
\hline
\end{tabular}
\end{table}

\clearpage
%

\clearpage

%

\clearpage

\clearpage


\end{document}